\documentclass[3p,preprint]{elsarticle} 
\usepackage{amssymb}
\usepackage{amsmath}
\usepackage{booktabs}
\usepackage{caption}
\usepackage{graphicx}
\PassOptionsToPackage{hyphens}{url}\usepackage[colorlinks=true]{hyperref}
\usepackage[utf8]{inputenc}
\usepackage[separate-uncertainty=true]{siunitx}

\biboptions{numbers,sort&compress}

\DeclareSIUnit{\patterns}{patterns}
\begin{document}

\title{Orientation dependent pinning of (sub)grains by dispersoids during recovery and recrystallization in an Al-Mn alloy}
\date{\today}

\author[1]{Håkon W. Ånes}
\author[2]{Antonius T. J. van Helvoort}
\author[1]{Knut Marthinsen\corref{corr_author}}
\ead{knut.marthinsen@ntnu.no}

\cortext[corr_author]{Corresponding author.}
\address[1]{Department of Materials Science and Engineering, Norwegian University of Science and Technology, 7491, Trondheim, Norway}
\address[2]{Department of Physics, Norwegian University of Science and Technology, 7491, Trondheim, Norway}

\journal{arXiv}


\begin{abstract}

The recrystallized grain size and texture in alloys can be controlled via the microchemistry state during thermomechanical processing.
The influence of concurrent precipitation on recovery and recrystallization is here analyzed by directly correlating (sub)grains of P, CubeND or Cube orientation with second-phase particles in a cold-rolled and non-isothermally annealed Al-Mn alloy.
The recrystallized state is dominated by coarse elongated grains with a strong P, weaker CubeND and even weaker Cube texture.
The correlated data enables orientation dependent quantification of the density and size of dispersoids on sub-boundaries and subgrains in the deformation zones around large constituent particles.
A new modified expression for the Smith-Zener drag from dispersoids on sub-boundaries is derived and used.
The results show that the drag on (sub)grain boundaries from dispersoids is orientation dependent, with Cube subgrains experiencing the highest drag after recovery and partial recrystallization.
The often observed size advantage of Cube subgrains in Al alloys is not realized due to the increased drag, thereby promoting particle-stimulated nucleation (PSN).
Relatively fewer and larger dispersoids in deformation zones around large particles give a reduced Smith-Zener drag on PSN nuclei, thus further strengthening the effect of PSN.
Observations substantiating the stronger P texture compared to the CubeND texture are a higher frequency of P subgrains and a faster growth of these subgrains.
The applied methodology enables a better understanding of the mechanisms behind the orientation dependent nucleation and growth behavior during recovery and recrystallization with strong concurrent precipitation in Al-Mn alloys.
In particular, the methodology gives new insights into the strong P and CubeND textures compared to the Cube texture.

\end{abstract}




\maketitle


\section{Introduction}

Al-Mn AA3xxx alloys are widely used in a broad range of products including beverage cans, automotive heat exchangers and packaging, where tailoring the grain size and texture during thermomechanical processing is key to achieve desired properties \cite{hirsch2013superior}.
Second-phase particles usually form in these alloys during casting, while a high concentration of alloying elements might remain in solid solution and precipitate during subsequent heat treatments.
During back annealing after hot and cold rolling, large constituent particles ($\gtrsim$ \SI{1}{\micro\meter}) may promote recrystallization by particle-stimulated nucleation (PSN) \cite{humphreys1977nucleation,engler1996nucleation}, while closely spaced fine dispersoids (sub-micron) may retard or even inhibit the transformation by pinning (sub)grain boundaries via Smith-Zener pinning \cite{rohrer2010introduction,humphreys2017recrystallization,huang2018doubleedge}.
When precipitated prior to or during recovery and recrystallization, the latter termed `concurrent precipitation', these dispersoids often result in slow recrystallization kinetics and coarse elongated grains \cite{vatne1997influence,somerday2003recrystallisation,tangen2010effect,huang2014influence,zhao2016orientation,huang2017controlling}.
The final recrystallization texture after strong PSN is usually weak, but with concurrent precipitation, unusually sharp P (typically \{011\}$\langle\bar{5}\bar{6}6\rangle$) and ND-rotated cube (\{001\}$\langle310\rangle$, CubeND) textures have been observed \cite{tangen2010effect,huang2014influence,zhao2016orientation,huang2017controlling}.
This strong P texture in particular is generally attributed to P (sub)grain boundaries experiencing a reduced Smith-Zener drag compared to those of CubeND and Cube (\{001\}$\langle100\rangle$) (sub)grains.
This assumption is based on extensive but indirect observations, where typically the microchemistry, in terms of number density and particle size, and the microstructure, in terms of grain size and texture, are analyzed separately.

Recently, the present authors demonstrated an experimental procedure to directly correlate (sub)grains and particles in a recovered Al-Mn alloy \cite{aanes2022correlated}.
The procedure enables correlated analysis of subgrains in deformation zones surrounding constituent particles and dispersoids located at (sub)grain boundaries.
In this work, this procedure is used to study the effect of large constituent particles and fine dispersoid particles on (sub)grains from the as deformed state to partial recrystallization with concurrent precipitation in a cold-rolled Al-Mn alloy.
The investigated alloy is the same as the one studied by Huang \textit{et al.} \cite{huang2014influence,huang2017controlling} (termed `C1' by them), in which they observed the unusually sharp P texture upon non-isothermal annealing with significant concurrent precipitation.
By using the same thermomechanical processing as them from deformation to full recrystallization, the objective of this work is to quantify the orientation dependent effects of particles on (sub)grains during recovery and recrystallization with concurrent precipitation.
It is demonstrated that these data can provide insight into the mechanisms behind the strong P and CubeND textures compared to the Cube texture with concurrent precipitation in Al-Mn alloys.


\section{Experimental methods}
\label{sec:experimental-methods}

The Al-Mn alloy supplied by Hydro Aluminium is direct-chill cast and has a composition of Al--0.53Fe--0.39Mn--0.152Si wt.\%.
It is homogenized in an air-circulation furnace from room temperature to \SI{450}{\celsius} at a heating rate of \SI{50}{\celsius\per\hour}, held for \SI{4}{\hour} and then water quenched.
The homogenized material is cold-rolled to a true strain of $\varepsilon = -\ln(h/h_0) = 3$, corresponding to a reduction in thickness $h$ of 95\%.
The material is finally annealed in an air circulation furnace at a heating rate of \SI{50}{\celsius\per\hour}, with samples retrieved at intervals of \SI{50}{\celsius} up to \SI{400}{\celsius}.

The recrystallization kinetics are characterized by measuring the electrical conductivity, hardness and macrotexture of the samples including the as deformed material.
These measurements are performed on the sample surface spanned by the rolling direction (RD) and the transverse direction (TD).
The hardness HV is measured with a \textit{Leica Pros VMHT Vickers} indenter with a load of \SI{1}{\kilogram} and a dwell time of \SI{15}{\second}, and the hardness per sample is an average of ten measurements with an uncertainty of two standard deviations.
The electrical conductivity is measured with a \textit{Foerster Sigmatest 2.069} tester, and the conductivity per sample is an average of 30 measurements with an uncertainty of two standard deviations.
The macrotexture is measured with X-ray diffraction (XRD) using a \textit{Siemens Diffractometer D5000} with a copper anode operated at \SI{40}{\kilo\volt} and a measurement spot with a diameter of about \SI{2}{\milli\meter}.
Samples are prepared for macrotexture measurements by mechanical polishing, etching in \textit{Alubeis} (a mixture of 10-20\% NaOH and sugar) and final etching in 20-30\% HNO$_3$.
Four incomplete pole figures \{111\}, \{200\}, \{220\} and \{311\} of Al, after correction with respect to background intensity and defocusing error, are used to calculate the orientation distribution function (ODF) using \textit{MTEX} v5.8 \cite{hielscher2008novel}.
Volume fractions, of typical texture components in cold-rolled and recrystallized Al, are then calculated using an orientation spread of \SI{15}{\degree} from the ideal orientation and an orthorhombic sample symmetry valid for cold-rolled samples.
The typical texture components in cold-rolled Al are copper (\{112\}$\left<11\bar{1}\right>$, C), S (\{123\}$\left<634\right>$) and brass (\{011\}$\left<2\bar{1}1\right>$, B), which make up the $\beta$-fiber \cite{humphreys2017recrystallization}.
The recrystallization texture components of interest in this work are the mentioned Cube, CubeND and P.
The total volume fraction can be higher than 100\% since the \SI{15}{\degree} spread leads to overlapping regions in orientation space for some texture components.

SEM measurements are performed on the sample surface spanned by the RD and the normal direction (ND).
The sample embedded in epoxy is mechanically polished using diamond paste down to \SI{1}{\micro\meter} grain size, followed by vibration polishing in a \textit{Buehler VibroMet 2} using colloidal silica of \SI{0.05}{\micro\meter} grain size.
The final deformed surface layer is removed with ion polishing in a \textit{Hitachi IM-3000} using Ar$^+$ ions accelerated to \SI{3}{\kilo\electronvolt} with the sample tilted \SI{60}{\degree} from the horizontal and rotating.
The sample is plasma cleaned with a \textit{Fischione 1020} prior to insertion in the SEM.
This sample preparation was chosen because it provides a sample surface sufficiently free of deformation for EBSD analysis, while only a limited number of particles are visibly removed from it.

Particles are detected in backscatter electron (BSE) images with a resolution of about \SI{0.026}{\micro\meter}, acquired in a \textit{Zeiss Ultra 55} FEG SEM operated at \SI{5}{\kilo\volt}.
A particle detection procedure for the sample at \SI{300}{\celsius} is described by the authors in Ref. \cite{aanes2022correlated}.
In order to achieve satisfactory quantitative particle detection from BSE images with varying intensity distributions, some modifications to this procedure are required.
To obtain comparable intensity distributions, intensities in all images are matched using histogram matching to the intensities in a BSE image obtained at \SI{300}{\celsius}.
Furthermore, instead of detecting particles via region-based segmentation, large and small particles are detected in two separate images using automatic thresholding.
Larger particles are detected in the histogram matched image after applying rolling ball averaging with a ball radius of one pixel, effectively blurring out the small particles.
Smaller particles are detected in the image obtained from subtracting the rolling ball averaged image from the histogram matched image, effectively masking out the larger particles.
After thresholding, these two images are combined into the final binary particle map, where any incorrectly detected particles are removed as described in Ref. \cite{aanes2022correlated}.
Image processing is done with \textit{ImageJ} \cite{schneider2012nih}.

(Sub)grain orientations and sizes are mapped using electron backscatter diffraction (EBSD) from the same regions of interest (ROIs) as the BSE images.
The same SEM as for particle detection is used, but operated at \SI{17}{\kilo\volt} with a sample tilt of \SI{70}{\degree} and a nominal step size of \SI{0.1}{\micro\meter}.
EBSD patterns are acquired with a \textit{NORDIF UF-1100} detector with a binning factor of 6, 8 or 9 to pattern resolutions of (96 $\times$ 96), (80 $\times$ 80) or (60 $\times$ 60) pixels, respectively, with acquisition speeds in the range of 70-350 patterns \si{\per\second}.
These acquisition parameters are chosen to provide a sufficient signal-to-noise ratio on the detector within a reasonable acquisition time.
Patterns are indexed as Al (space group $Fm\bar{3}m$, $a$ = \SI{0.404}{\nano\meter}) by dictionary indexing \cite{chen2015dictionary} followed by refinement as implemented in \textit{kikuchipy} v0.6 \cite{pena2017electron,kikuchipy_0_6_1}, using a dynamical master pattern simulated with \textit{EMsoft} v4.3 \cite{callahan2013dynamical}.
An average projection center (PC) for each dataset is obtained by Hough indexing of (240 $\times$ 240) pixel calibration patterns with \textit{PyEBSDIndex} v0.1 \cite{pyebsdindex}.

\begin{figure}[htbp]
  \centering
  \includegraphics[width=0.5\columnwidth]{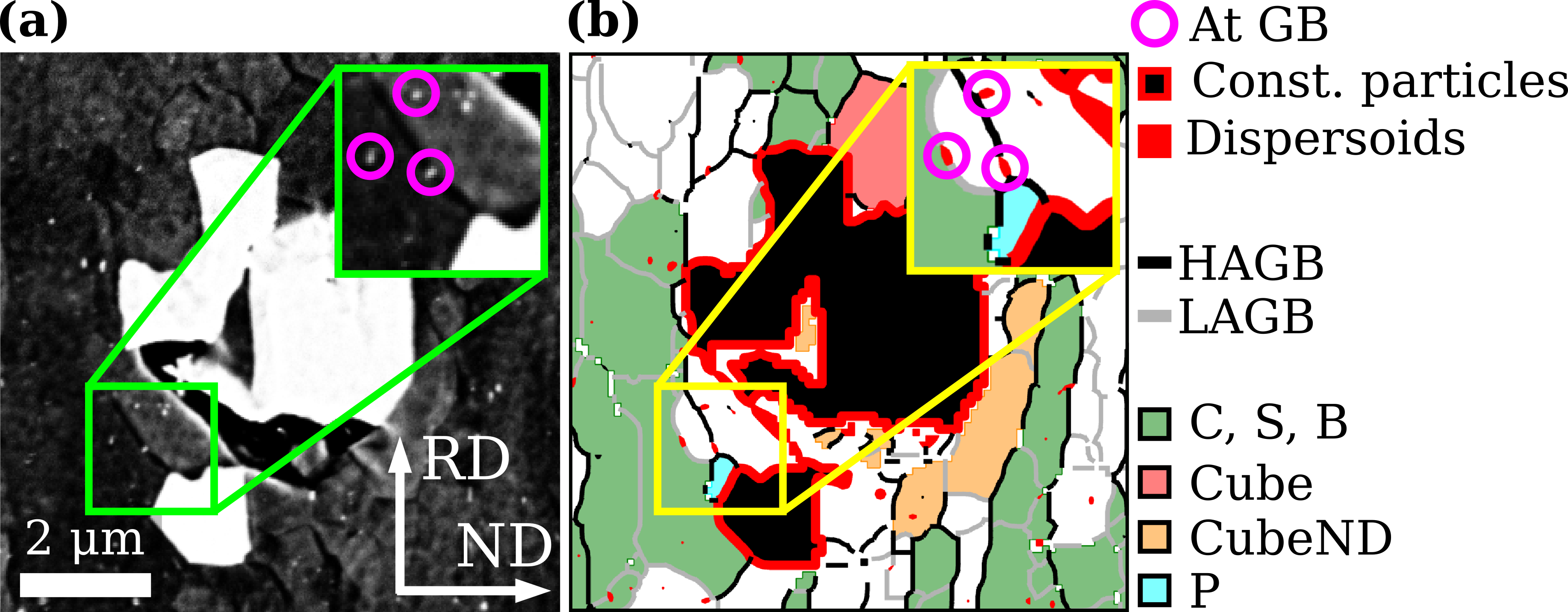}
  \caption{
  An example of a small region of a multimodal dataset obtained following a procedure similar to the one reported in Ref. \cite{aanes2022correlated}, where detected particles in a BSE image (a) are marked in an EBSD map (b) which is corrected for distortions.
  The multimodal dataset contains (sub)grains classified by texture component, low-angle or high-angle boundaries (LAGB or HAGB) between subgrains, larger constituent particles and smaller dispersoid particles, some of which are located at boundaries as highlighted in the inset.
  }
  \label{fig:multimodal-dataset}
\end{figure}

In order to analyze particles and (sub)grains from the same ROI in the same reference frame, the EBSD and particle maps are overlapped via image registration as described in Ref. \cite{aanes2022correlated}, accounting for the different step sizes and any distortions in the EBSD map.
Three multimodal datasets are thus obtained from each condition of interest, covering a total area of 0.020--0.028 \si{\milli\meter\squared} per condition.
Orientations are analyzed with \textit{MTEX} \cite{bachmann2011grain} as detailed in Ref. \cite{aanes2022correlated}.
No averaging or clean-up of orientations is required even in the datasets of the as deformed state, as dictionary indexing is capable of indexing EBSD patterns with a relatively low signal-to-noise ratio \cite{singh2018high}.
The particle map is used to create a `dual phase' dataset to distinguish between points with Al and points with particles.
Reconstructed (sub)grains are classified by texture components based on the lowest misorientation angle between the (sub)grain mean orientation and the components' ideal orientations within a misorientation angle threshold of \SI{15}{\degree}.
(Sub)grains outside the threshold are considered `random'.
As with the macrotexture measurements, an orthorhombic sample symmetry is assumed when classifying (sub)grains.
(Sub)grain sizes $D$ and particle sizes $d$ are given as equivalent circular diameters $0.816 \cdot 2 \cdot \sqrt{A/\pi}$, where $A$ is the (sub)grain or particle area.
The prefactor follows from considering that the observed 2D plane cuts 3D particles at random \cite{humphreys2017recrystallization}.
An example of a small part of a BSE image and the resulting multimodal dataset are shown in Fig. \ref{fig:multimodal-dataset} (a) and (b) respectively, where particles possibly smaller than the EBSD step size are included in the same dataset as (sub)grains and boundaries.
Overviews of all multimodal datasets used in this paper are given in Figs. \ref{fig:maps-0}-\ref{fig:maps-325c} in the supplementary material.

\section{Results}


\subsection{Recrystallization kinetics}

\begin{figure}[htbp]
  \centering
  \includegraphics[width=0.5\columnwidth]{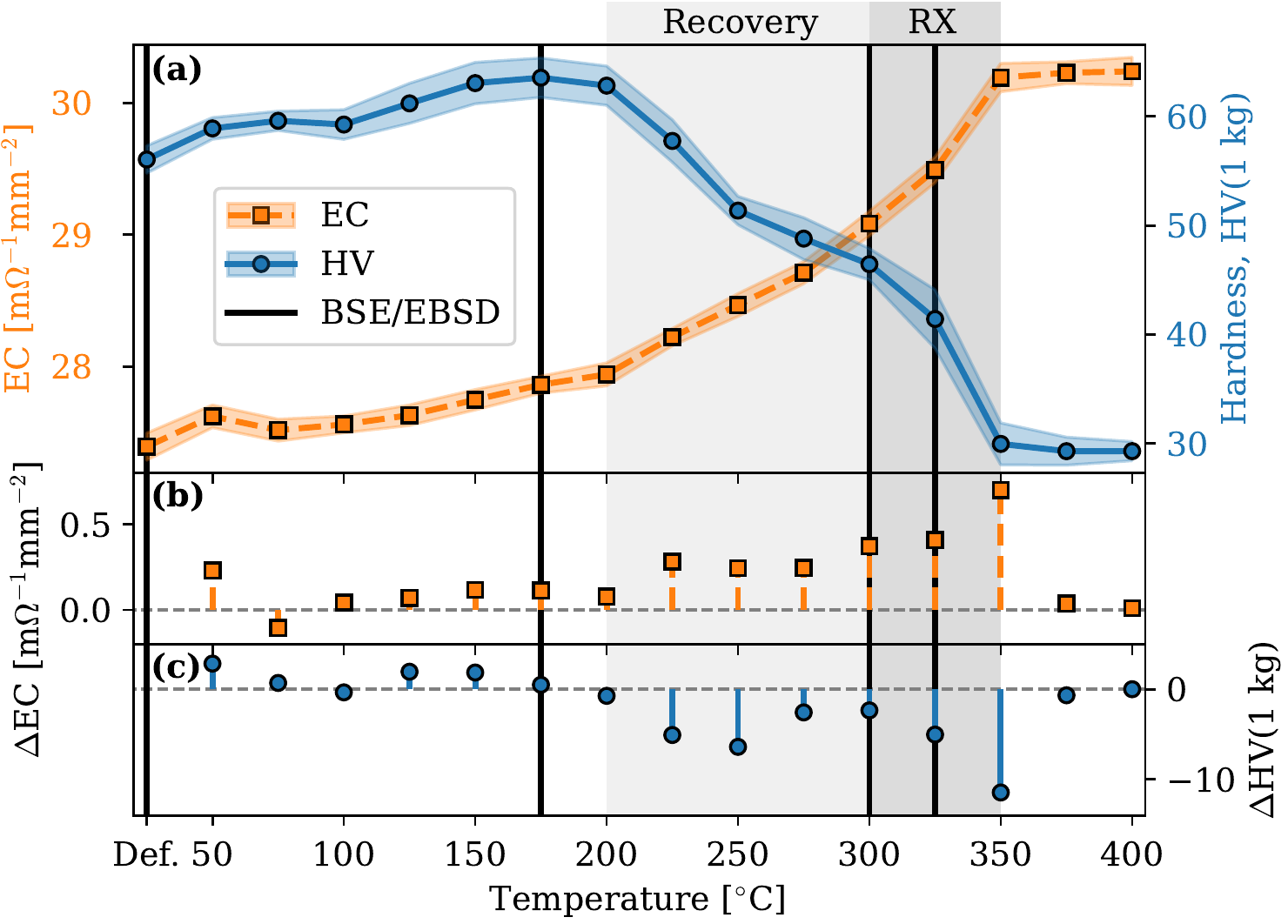}
  \caption{
  Recrystallization kinetics of the cold-rolled and non-isothermally annealed Al-Mn alloy.
  (a) Electrical conductivity (EC, left) and hardness (right) and their rate of change in (b) and (c), respectively.
  The temperature intervals for recovery and recrystallization (RX) are highlighted in light and darker gray, respectively.
  Four conditions selected for detailed analysis of particle and subgrain statistics are highlighted in bold vertical lines.
  }
  \label{fig:rx-kinetics}
\end{figure}

The changes in electrical conductivity (EC) and hardness as a function of the non-isothermal annealing temperature are shown in Fig. \ref{fig:rx-kinetics} (a) with the rate of change in (b, c).
The material starts to recover above \SI{200}{\celsius} and is fully recrystallized above \SI{350}{\celsius}, at which point it has a coarse elongated grain structure.
BSE imaging and EBSD show no clearly recrystallized grains at \SI{300}{\celsius}, but that the material is partly recrystallized at \SI{325}{\celsius}.
This indicates that recrystallization starts just above \SI{300}{\celsius}.
The general increase in conductivity with annealing is evidence of precipitation concurrent with recovery and recrystallization, with an increased rate during recovery from \SIrange{200}{300}{\celsius} and an even higher rate during recrystallization from \SIrange{300}{350}{\celsius}.
Minimal precipitation is observed after recrystallization has completed.
To check the possibility for further precipitation, the conductivity is measured of the material non-isothermally annealed to \SI{400}{\celsius}, and then held at that temperature for \SI{e5}{\second}.
The conductivity after this holding time increased by \SI{1.23(4)}{\meter\per\ohm\per\milli\meter\squared}, i.e. some further precipitation occurred.
This shows that what is practically a stop in precipitation after complete recrystallization is temporary, and that further annealing at higher temperatures for longer times is required to continue the precipitation.
A similar behavior was observed by Vatne \textit{et al.} \cite{vatne1997influence} in an AA3103 alloy where precipitation occurred primarily during recrystallization, irrespective of the isothermal annealing temperature.
They argued that this indicated that precipitation was not diffusion controlled, but that elements in solid solution precipitated on the moving boundaries of recrystallized grains, a description which fits the precipitation behavior observed here as well.

\begin{figure}[htbp]
  \centering
  \includegraphics[width=0.5\columnwidth]{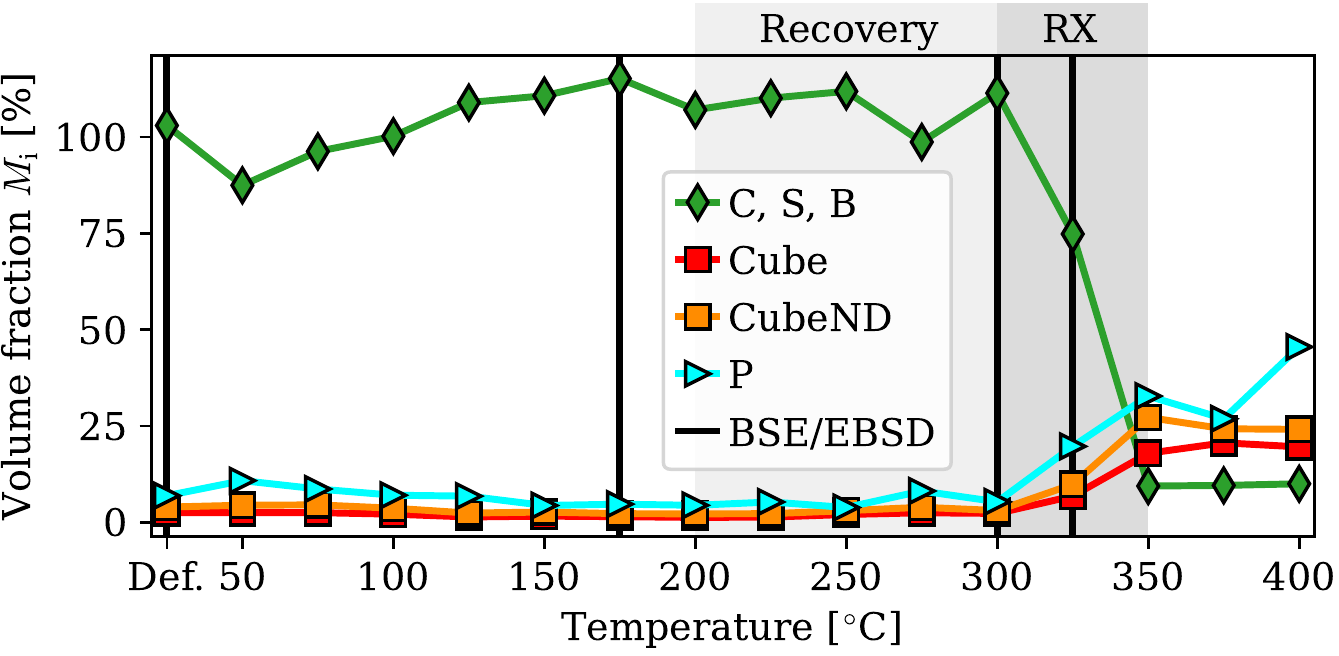}
  \caption{
  XRD macrotexture volume fractions $M_{\mathrm{i}}$ per texture component in the cold-rolled and non-isothermally annealed Al-Mn alloy.
  Volume fractions are given as a function of annealing temperature from the as deformed (`Def.') to after the fully recrystallized state.
  The temperature intervals for recovery and recrystallization (RX) are highlighted in light and darker gray, respectively.
  Four conditions selected for detailed analysis of particle and subgrain statistics are highlighted in bold vertical lines.
  }
  \label{fig:volume-fractions}
\end{figure}

The XRD macrotexture volume fractions $M_{\mathrm{i}}$ per texture component are plotted as a function of annealing temperature in Fig. \ref{fig:volume-fractions}.
As expected for cold-rolled Al, the as deformed microstructure is characterized by a strong $\beta$-fiber texture running from the C component over S to B.
Upon recrystallization, this deformation texture is superseded by a recrystallization texture dominated by the P, CubeND and Cube components.
The ratios of P to CubeND and Cube volume fractions after recrystallization at \SI{400}{\celsius} are about 1.9 and 2.3, respectively.

The as deformed state, softening and precipitation behavior and recrystallized grain structure and texture are all comparable to those observed by Huang \textit{et al.} \cite{huang2014influence,huang2017controlling} for the same alloy and thermomechanical processing.
Their final recrystallization texture was characterized after non-isothermal annealing to \SI{400}{\celsius} followed by an additional holding time of \SI{e5}{\second} at the same temperature.
The P texture was reported to be about 4 and 6.7 times stronger than CubeND and Cube, respectively.
The relatively stronger P texture reported by them compared to the present strength is assumed to be a result of grain coarsening following primary recrystallization.
These comparisons show that recovery and recrystallization in the alloy in this work is similar to that reported by Huang \textit{et al.}.
This allows us to draw from their characterization of the recrystallized material, and instead focus on the causes of the dominating P texture after recovery and recrystallization with significant concurrent precipitation.
To that end, four conditions of interest are selected for detailed analysis of particles and (sub)grain statistics, namely the as deformed state and the material before recovery at \SI{175}{\celsius}, after recovery at \SI{300}{\celsius} and after partial recrystallization at \SI{325}{\celsius}.
These conditions are highlighted by bold vertical lines in Figs. \ref{fig:rx-kinetics} and \ref{fig:volume-fractions}.


\subsection{Precipitation kinetics}

\begin{figure*}[htbp]
  \centering
  \includegraphics[width=\textwidth]{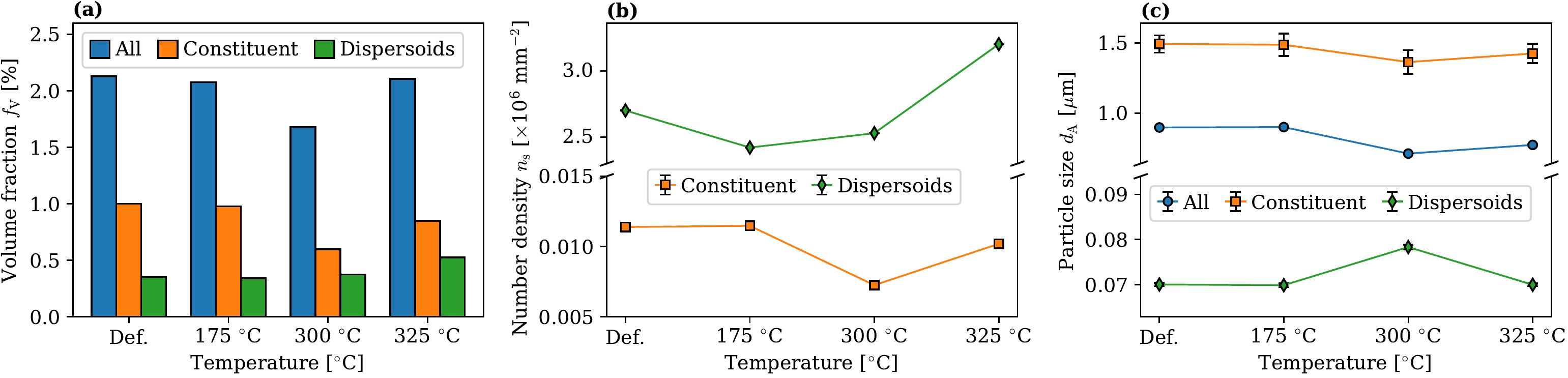}
  \caption{
  Precipitation kinetics of the cold-rolled and non-isothermally annealed Al-Mn alloy.
  (a) Volume fraction $f_{\mathrm{V}}$, (b) number density $n_{\mathrm{s}}$ with 95\% confidence intervals and (c) area weighted particle size $d_{\mathrm{A}}$ with 95\% confidence intervals as a function of annealing temperature from the as deformed (`Def.') to the partly recrystallized state for all, constituent and dispersoid particles separately.
  In (b), the number density of all particles is omitted, and the confidence intervals are so narrow they do not show.
  Note the broken vertical axes in (b, c).
  }
  \label{fig:particle-stats}
\end{figure*}

Before correlating particle locations with (sub)grains and (sub)grain boundaries, the precipitation kinetics and subgrain growth from the four analyzed conditions are presented separately.
The changes in particle volume fraction $f_{\mathrm{V}}$ as estimated from the surface area fraction, the number of particles per area $n_{\mathrm{s}}$ and the area weighted particle size $d_{\mathrm{A}}$ from the as deformed to the partly recrystallized state are shown in Fig. \ref{fig:particle-stats}.
Values are presented for all particles, constituent particles and dispersoid particles.
Particles with sizes $d$ in the range 0.03--0.24 \si{\micro\meter} are classified as dispersoids based on the approximately bimodal particle size distributions in all conditions (Fig. \ref{fig:particle-size-hist} in the supplementary material).
The critical lower diameter of particles considered to be potential nucleation sites for recrystallized grains, i.e. contributing to PSN and in the following labeled `constituent', after rolling to 95\% is about \SI{1}{\micro\meter} \cite{humphreys1977nucleation}.
The lower diameter threshold for constituent particles used here is \SI{0.8}{\micro\meter} due to the stereological considerations mentioned at the end of \S\ref{sec:experimental-methods}.
In each condition, the total number of particles, constituent particles and dispersoids ranges from about \SIrange{68000}{118000}{}, \SIrange{150}{300}{} and \SIrange{45000}{71000}{} for the three populations, respectively.
The volume fraction $f_{\mathrm{V}}$ of all particles is about 2\%, and is approximately the same for all particles and constituent particles in the as deformed and the partly recrystallized state.
The drop in $f_{\mathrm{V}}$ at \SI{300}{\celsius} seen in Fig. \ref{fig:particle-stats} results from fewer detected constituent particles, which is most likely due to their partial inhomogeneous dispersion in the deformed material.
The number density $n_{\mathrm{s}}$ and size of constituent particles are approximately constant during recovery and recrystallization (gray ranges in Figs. \ref{fig:rx-kinetics} and \ref{fig:volume-fractions}).
The volume fraction of dispersoids increases during recovery and substantially during recrystallization, in line with increases in the rate of change of conductivity in Fig. \ref{fig:rx-kinetics} (b).
The increase in $f_{\mathrm{V}}$ during recovery results from coarsening of dispersoids, giving an increased average dispersoid size.
The substantial increase in $f_{\mathrm{V}}$ during recrystallization results mainly from increased precipitation, as evident from the increased $n_{\mathrm{s}}$.
The increased $n_{\mathrm{s}}$ during recrystallization corresponds to a reduced average dispersoid size.

To assess whether PSN nuclei experience a reduced Smith-Zener drag from dispersoids in the initial stages of recrystallization, the number of dispersoids per area $n_{\mathrm{s}}$ in an area a certain distance from constituent particles, often called the dispersoid-free zone \cite{furrer1979recrystallization,tangen2010effect}, relative to $n_{\mathrm{s}}$ elsewhere is calculated.
The relative density in the area extending \SI{0.5}{\micro\meter} out from constituent particles is about 95\% before recovery and drops to about 80\% after partial recrystallization.
When extending the area out to \SI{1}{\micro\meter}, there is no clear difference in $n_{\mathrm{s}}$ close to constituent particles and elsewhere.
Furthermore, dispersoids in the area out to \SI{0.5}{\micro\meter} have a diameter about 1.5 times greater than dispersoids elsewhere.
These results indicate that while the area within \SI{0.5}{\micro\meter} of constituent particles is not free of dispersoids, subgrains in this area experience a reduced Smith-Zener drag upon growth since they encounter fewer and larger dispersoids than subgrains elsewhere.


\subsection{Subgrain growth}

\begin{figure*}[htbp]
  \centering
  \includegraphics[width=\textwidth]{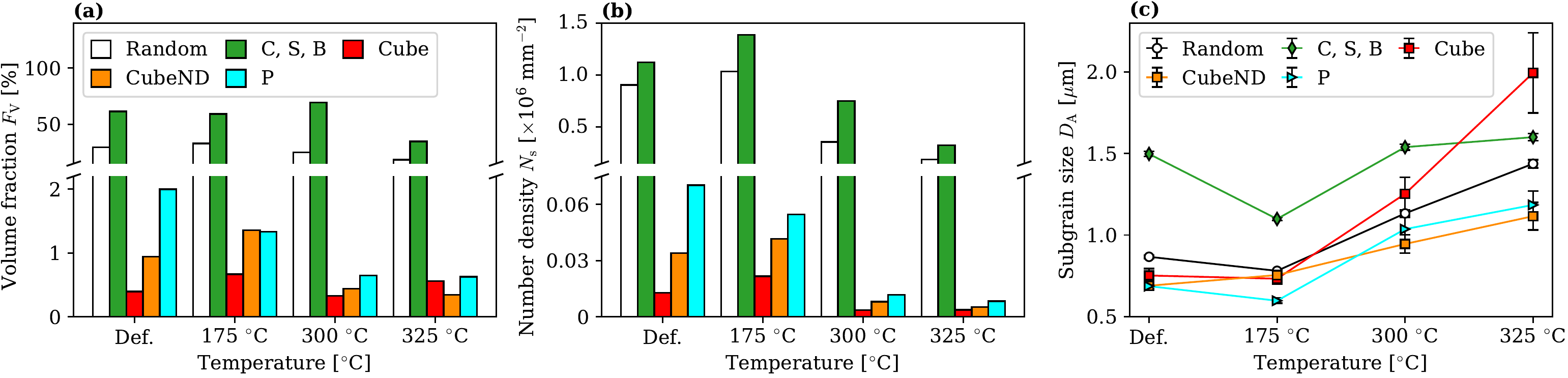}
  \caption{
  Subgrain statistics in the cold-rolled and non-isothermally annealed Al-Mn alloy.
  Values are shown per texture component from the as deformed (`Def.') to the partly recrystallized state.
  (a) Volume fraction $F_{\mathrm{V}}$, (b) number of subgrains per area and (c) area weighted subgrain size $D_{\mathrm{A}}$ with 95\% confidence intervals.
  Note the broken vertical axes in (a, b).
  }
  \label{fig:subgrain-stats}
\end{figure*}

Changes in subgrain volume fraction $F_{\mathrm{V}}$, number of subgrains per area $N_{\mathrm{s}}$ and the area weighted subgrain size $D_{\mathrm{A}}$ from the as deformed to the partly recrystallized state are shown in Fig. \ref{fig:subgrain-stats} per texture component and for all other subgrains, labeled `random' subgrains.
In each of the four conditions, the total number of subgrains analyzed ranges from about \SI{60000}{} in the as deformed state to about \SI{10900}{} in the partly recrystallized state.
Recrystallized grains in the partly recrystallized state are excluded from this analysis.
A recrystallized grain is classified as having a grain size $D >$ \SI{4}{\micro\meter}, a grain orientation spread $<$ \SI{1}{\degree} and with $>$ 50\% of its boundary having a misorientation angle $\omega >$ \SI{15}{\degree}, which is considered a high angle grain boundary (HAGB).
In the as deformed state, there are mostly subgrains of deformation orientations, labeled `deformation' subgrains, and random subgrains.
Both the volume fraction $F_{\mathrm{V}}$ and number density $N_{\mathrm{s}}$ of P subgrains are two and five times higher than that of CubeND and Cube subgrains, respectively.
This indicates that the P texture has a frequency advantage.
All subgrains are of a similar size [Fig. \ref{fig:subgrain-stats} (c)], except for the deformation subgrains which are much larger.
This is due to their location in deformation bands of similar orientations extending in the rolling direction.
With annealing, the volume fractions of deformation and random subgrains stay approximately constant, while their number densities decrease after recovery and partial recrystallization.
The volume fractions of subgrains of recrystallization orientations, labeled `recrystallization' subgrains, are all between 0.5--2\% until partial recrystallization.
While their volume fractions are comparable at this condition, the number density of P subgrains remain higher than the number densities of CubeND and Cube subgrains.
All subgrains apart from deformation subgrains grow in size during recovery and recrystallization.
The P and CubeND subgrains are the smallest in the partly recrystallized state.

\subsection{Effects of constituent particles on subgrains}
\label{sec:effects-constituent-particles-subgrains}

Since the particles and subgrains making up the above statistics in Figs. \ref{fig:particle-stats} and \ref{fig:subgrain-stats} are acquired within the same regions of interest, they can be combined into multimodal datasets \cite{aanes2022correlated}.
Each map point in the datasets contains parts of a particle and its size, including those smaller than the EBSD step size, or an Al orientation if there is no particle there.
The location and area of effect of larger particles are more challenging to characterize in spatially correlated 2D maps compared to smaller particles.
This is because it is possible that a large particle is located beneath or was located above the polished surface, and thus might have affected the (sub)grains on the analyzed surface.
Zhang \textit{et al.} discussed this scenario when they spatially correlated recrystallization nuclei and particles in a multimodal 3D dataset from an Al-Mn alloy cold-rolled and annealed to partial recrystallization \cite{zhang2012three}.
They observed that the fraction of nuclei at clusters or bands of large particles was underestimated from a single 2D slice compared to the fraction from the full 3D dataset.
To avoid that this underestimation affects the analysis in this work, the potential of PSN is compared between subgrain populations only and not to effects of other nucleation mechanisms.

\begin{figure*}[htbp]
  \centering
  \includegraphics[width=\textwidth]{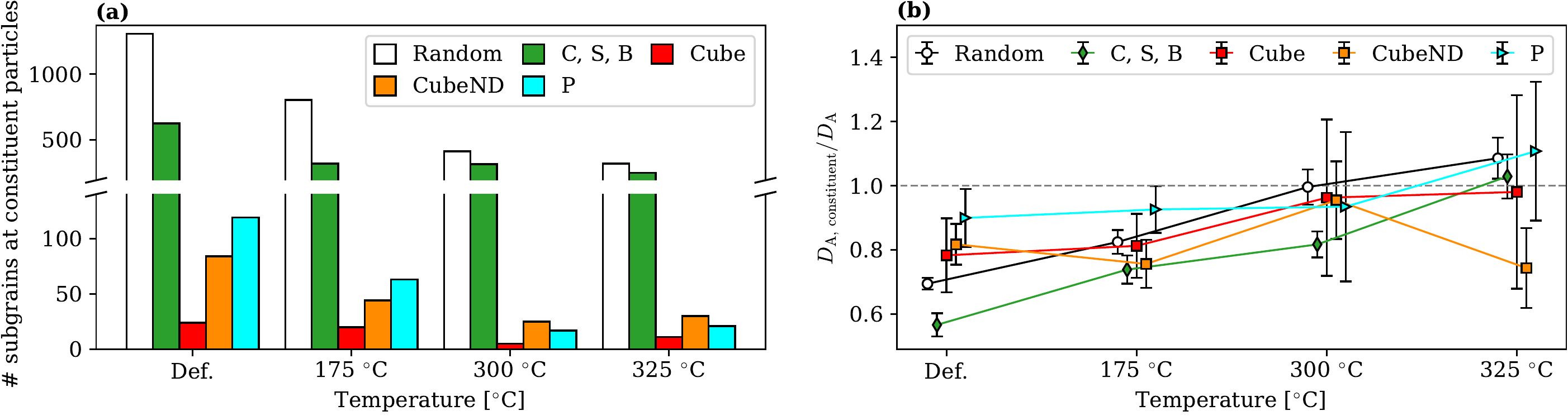}
  \caption{
  Subgrains at constituent particles in the cold-rolled and non-isothermally annealed Al-Mn alloy.
  Values are shown per texture component from the as deformed (`Def.') to the partly recrystallized state.
  (a) Number of subgrains as a function of texture component.
  (b) Ratio of sizes $D_{\mathrm{A, constituent}}$ of subgrains at constituent particles over sizes $D_{\mathrm{A}}$ of subgrains elsewhere in the microstructure, with 95\% confidence intervals.
  Note the broken vertical axis in (a).
  Values in (b) are slightly shifted horizontally to show the individual confidence intervals.
  }
  \label{fig:subgrains-at-constituent-particles}
\end{figure*}

A subgrain is considered to be at a constituent particle if one or more of its neighboring map points is a constituent particle.
Changes in the number of subgrains at constituent particles per texture component, from the as deformed to the partly recrystallized state, are given in Fig. \ref{fig:subgrains-at-constituent-particles} (a).
There are more random subgrains than deformation subgrains here, even though there are more of the latter in the overall microstructure [Fig. \ref{fig:subgrain-stats} (a, b)].
A wide range of random subgrains in the deformation zones around large particles is common in cold-rolled Al alloys with such particles \cite{engler1996formation,humphreys2017recrystallization}.
This means that a strong PSN effect, without orientation dependent pinning of (sub)grain boundaries, usually result in a weak recrystallization texture.
This weak texture is a result of a large amount of randomly oriented grains and only a slight preference of CubeND and P grains \cite{engler1995texture,engler1996formation}, and certainly not one dominated by P and CubeND as observed in this alloy both here and in previous works \cite{huang2017controlling}.
In the as deformed state, subgrains at constituent particles are smaller than subgrains elsewhere in the microstructure, as evident from the ratio $D_{\mathrm{A,\:constituent}} / D_{\mathrm{A}}$ presented in Fig. \ref{fig:subgrains-at-constituent-particles} (b).
This is in accordance with a previous comparison of size distributions of subgrains close to and far away from constituent particles \cite{humphreys1977nucleation}.
The 95\% confidence intervals for the subgrain size ratios, calculated from error propagation, show relatively high uncertainties for the recrystallization subgrains after recovery.
Even so, the ratio increases close to 1 or above for all but CubeND subgrains after partial recrystallization, which means that most subgrains at constituent particles grow faster than subgrains elsewhere.
This is a clear indication of a growth advantage of subgrains at constituent particles.
This advantage can be understood in terms of both the increased stored energy in the deformation zone around these particles and the reduced dispersoid density and larger dispersoids in this zone.

\subsection{Effects of dispersoids on subgrains}

A dispersoid is considered to be at a (sub)grain boundary if it is within \SI{0.1}{\micro\meter} (the EBSD step size) of it.
The cumulative probability of each dispersoid's distance to the closest (sub-)boundary is given in Fig. \ref{fig:dispersoid-distance-to-gb} in the supplementary material.
About 60\% of dispersoids are located at (sub-)boundaries before recrystallization starts, while only about 40\% are after partial recrystallization.
If a boundary segment's area is estimated as its length times 2 $\times$ \SI{0.1}{\micro\meter}, which is an overestimation, boundaries make up about 50\% of the total area before recovery, 40\% after recovery and 30\% after partial recrystallization.
Dispersoids formed prior to final annealing in Al alloys are usually observed to be randomly distributed in the deformed microstructure \cite{humphreys2017recrystallization,huang2018doubleedge}, and this study finds the same.

\begin{figure}[htbp]
  \centering
  \includegraphics[width=0.5\columnwidth]{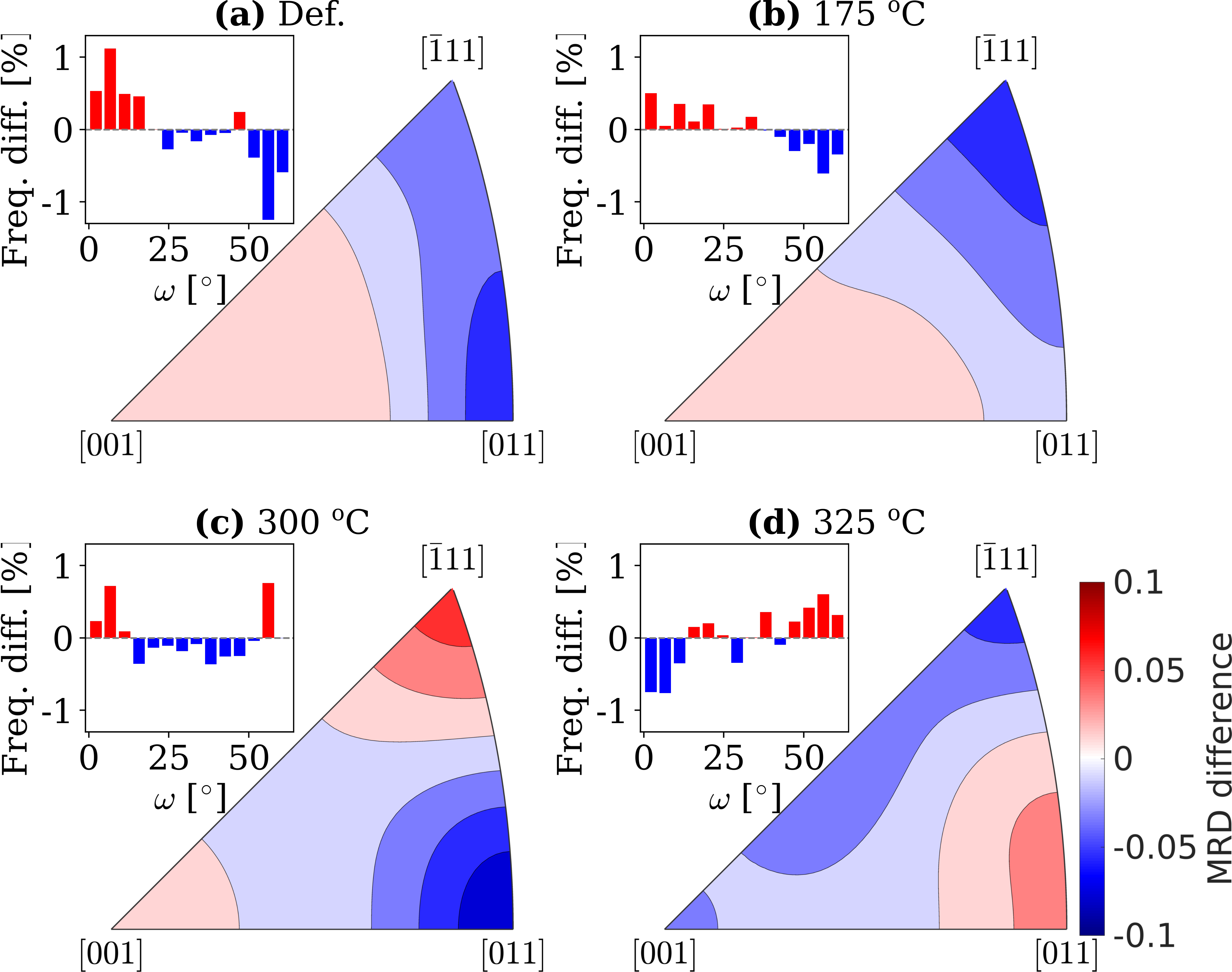}
  \caption{
  Dependence of dispersoid densities $f_{\mathrm{L}}$ on misorientation axis and angle, with higher and lower densities above (red) and below (blue) zero, respectively.
  Relative densities are calculated for the (a) as deformed (`Def') state, (b) before recovery at \SI{175}{\celsius}, (c) after recovery at \SI{300}{\celsius} and (d) after partial recrystallization at \SI{325}{\celsius} of the cold-rolled and non-isothermally annealed Al-Mn alloy.
  }
  \label{fig:dispersoids-misorientations}
\end{figure}

In Al-Mn alloys, precipitation during final annealing takes place preferentially at HAGBs rather than low-angle grain boundaries (LAGBs) or dislocations \cite{somerday2003recrystallisation,tangen2010effect}.
The dependence of the number of dispersoids per sub-boundary length $f_{\mathrm{L}}$ on the boundary misorientation axis and angle can be investigated with the present data.
To express the former, the misorientation axis distributions of all boundary segments and of segments with dispersoids on them are calculated.
By subtracting the former distribution from the latter, the relative dispersoid density $f_{\mathrm{L}}$ per misorientation axis is obtained.
The density is expressed as the difference in multiples of random density (MRD) away from zero, and visualized in the inverse pole figure of point group $m\bar{3}m$.
A misorientation axis having an MRD difference greater (lower) than zero means that the dispersoid density on boundaries with this axis is higher (lower) than on boundaries with other axes.
Likewise, the relative dispersoid density per misorientation angle is obtained by subtracting the angle probability histograms of all boundary segments and of segments with dispersoids on them.
This density is expressed as a frequency difference in percentage away from zero.
Fig. \ref{fig:dispersoids-misorientations} shows the dependence of the relative dispersoid density on sub-boundaries' misorientation axes and angles from the as deformed to the partly recrystallized state.
While variations in densities are small, there are some changes with annealing temperature worth noting.
In the as deformed state, there are more dispersoids on LAGBs and fewer on HAGBs.
With annealing, this difference disappears, until it is reversed in the partly recrystallized state when there are more dispersoids on HAGBs and fewer on LAGBs.
Differences in $f_{\mathrm{L}}$ on misorientation axes are seen after recovery and partial recrystallization, when there is a strong concurrent precipitation reaction.
After recovery, the relative densities are $f_{\mathrm{L,\left<111\right>}} > f_{\mathrm{L,\left<001\right>}} \gg f_{\mathrm{L,\left<011\right>}}$.
After partial recrystallization, the differences in densities have reduced and changed to $f_{\mathrm{L,\left<011\right>}} > f_{\mathrm{L,\left<001\right>}} > f_{\mathrm{L,\left<111\right>}}$.
Changes in the dependence of $f_{\mathrm{L}}$ on misorientation axes and angles are caused by both preferential precipitation of new dispersoids and increased pinning of moving boundaries.
The observation of an increased $f_{\mathrm{L}}$ on HAGBs with increasing precipitation during final annealing [Figs. \ref{fig:rx-kinetics} (a, b) and \ref{fig:particle-stats} (b)] is in line with previous studies \cite{somerday2003recrystallisation,tangen2010effect}.
The high overall number density $n_{\mathrm{s}}$ of dispersoids provides a high probability of pinning moving boundaries.
The initial increased $f_{\mathrm{L}}$ on $\left<111\right>$-boundaries after recovery can be understood in terms of these boundaries having an increased mobility \cite{lucke1974orientation,yang2001measuring,fan2014oriented} and thus sweeping more dispersoids upon growth [Fig. \ref{fig:subgrain-stats} (c)] compared to other boundaries.
That $\left<111\right>$-sub-boundaries then have a lower $f_{\mathrm{L}}$ after partial recrystallization seems to indicate that dispersoids precipitated on sub-boundaries during recrystallization favor those of other sub-boundaries than $\left<111\right>$-sub-boundaries.

\begin{figure}[htbp]
  \centering
  \includegraphics[width=0.5\columnwidth]{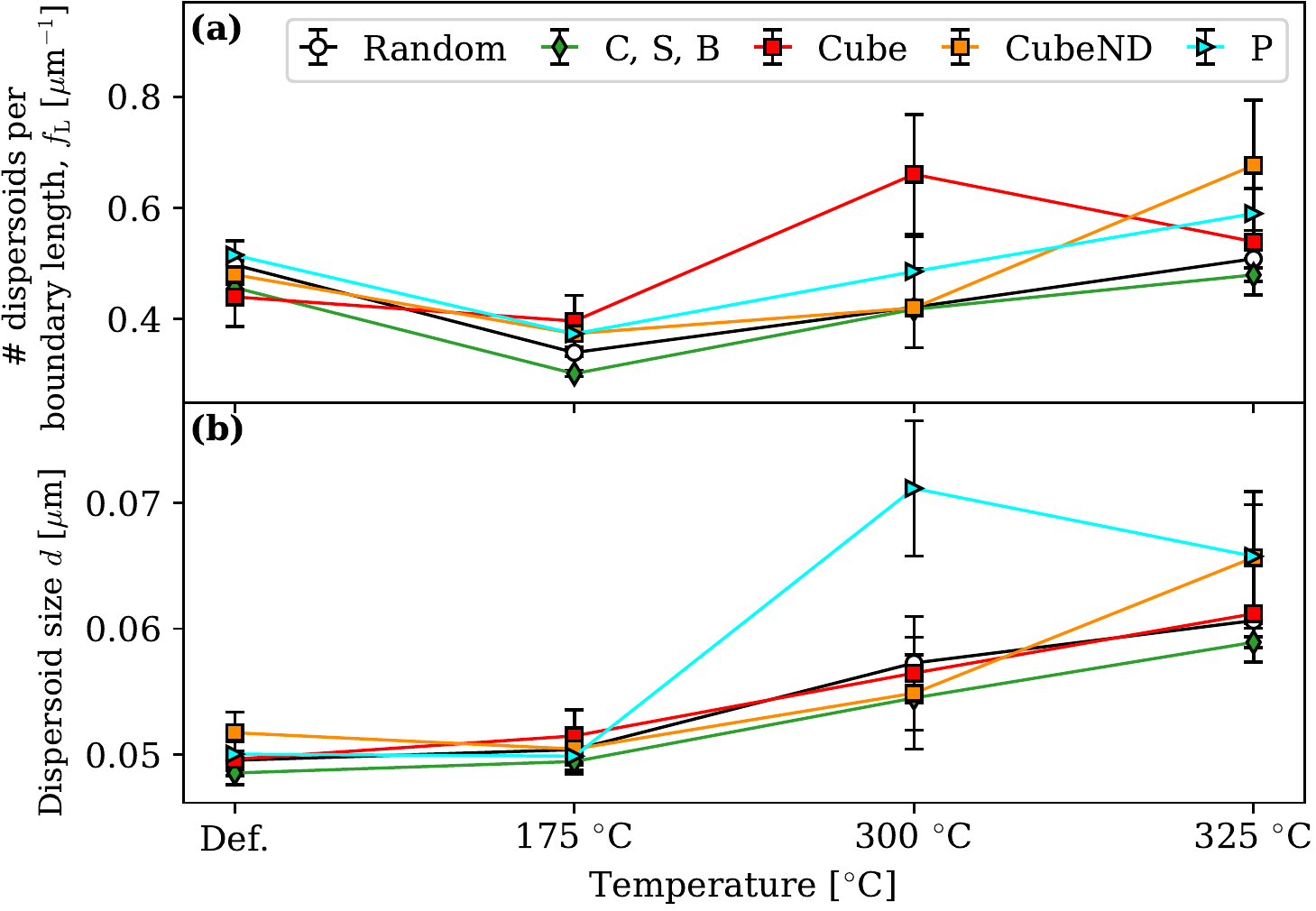}
  \caption{
  Dispersoids at sub-boundaries in the cold-rolled and non-isothermally annealed Al-Mn alloy.
  Values are shown per texture component from the as deformed (`Def.') to the partly recrystallized state, with 95\% confidence intervals.
  (a) Number of dispersoids per boundary length.
  (b) Dispersoid size per boundary length.
  }
  \label{fig:dispersoids-at-gb-per-comp}
\end{figure}

Variations in the dispersoid density $f_{\mathrm{L}}$ and the average dispersoid size $d$ on sub-boundaries can also be investigated as a function of texture component.
In Fig. \ref{fig:dispersoids-at-gb-per-comp} (a) and (b), the dispersoid density and size are shown from the as deformed to the partly recrystallized state.
These parameters are comparable for all boundaries before recovery, another indication that dispersoids are distributed quite randomly in the microstructure.
Cube and P subgrains grow at a higher rate than other subgrains during recovery [see Fig. \ref{fig:subgrain-stats} (c)], which coincides with those subgrains' boundaries having a higher $f_{\mathrm{L}}$ at this condition, with the highest density on Cube sub-boundaries.
The observations of increased $f_{\mathrm{L}}$ on $\left<111\right>$-sub-boundaries [Fig. \ref{fig:dispersoids-misorientations} (c)] and boundaries of fast growing subgrains are both indications that there is a positive correlation between subgrain growth and an increased sub-boundary dispersoid density during recovery.
It also supports the interpretation of the $\left<111\right>$-sub-boundaries being more mobile than other sub-boundaries during recovery.
As the dispersoids overall coarsen during recovery [Fig. \ref{fig:particle-stats} (c)], so do the dispersoids on all sub-boundaries.
P sub-boundaries stand out by accommodating much coarser dispersoids [Fig. \ref{fig:dispersoids-at-gb-per-comp} (b)].
After partial recrystallization, the overall number density of dispersoids $n_{\mathrm{s}}$ increases substantially.
However, only CubeND sub-boundaries experience a substantial increase in dispersoid density $f_{\mathrm{L}}$, while other sub-boundaries maintain a similar density as before recrystallization.
At this condition, all sub-boundaries except for P sub-boundaries experience a coarsening of dispersoids on them as well, with dispersoids on CubeND sub-boundaries coarsening the most.
These results can be combined to give an indication of the Smith-Zener drag on sub-boundaries.

\begin{figure}[htbp]
  \centering
  \includegraphics[width=0.5\columnwidth]{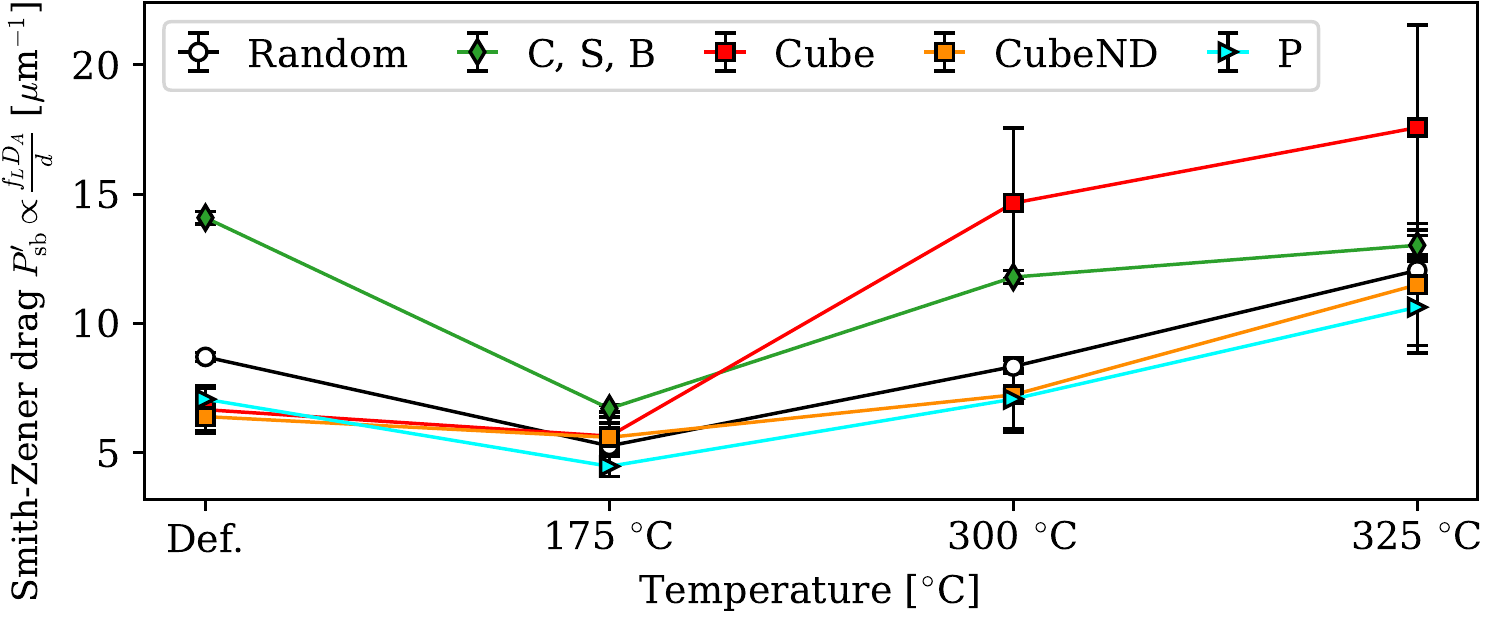}
  \caption{
  Modified Smith-Zener drag $P_{\mathrm{sb}}'$ at sub-boundaries in the cold-rolled and non-isothermally annealed Al-Mn alloy, calculated from Eq. \eqref{eq:smith-zener-drag}.
  Values are shown per texture component from the as deformed (`Def.') to the partly recrystallized state, with 95\% confidence intervals.
  }
  \label{fig:smith-zener-drag}
\end{figure}

The Smith-Zener drag from randomly distributed dispersoids of average radius $d/2$ on near-planar grain boundaries of average energy $\gamma$ is given by $P_{\mathrm{SZ}} = 3\gamma f_{\mathrm{V}}/d$ \cite{rohrer2010introduction,humphreys2017recrystallization}.
However, the fine subgrains and high number density of dispersoids $n_{\mathrm{s}}$ in the as deformed state provide a strong (sub)grain boundary-dispersoid correlation, as about 60\% of dispersoids are found at boundaries until the onset of recrystallization while about 40\% are at boundaries after partial recrystallization.
Furthermore, the interparticle spacing $n_{\mathrm{s}}^{-0.5}$ of dispersoids calculated from values in Fig. \ref{fig:particle-stats} (b) is in the order of half the average subgrain size $D_{\mathrm{A}}$, meaning that the assumption that dispersoids encounter a near-planar boundary is incorrect.
For these reasons, a modified expression for $P_{\mathrm{SZ}}$ must be used.
Hutchinson and Duggan \cite{hutchinson1978influence} argued that when a large fraction of the particles lie at sub-boundaries, the drag force per unit area of sub-boundary can be written as $P_{\mathrm{sb}} = 3\gamma_{\mathrm{sb}}f_{\mathrm{V}}/(A_{\mathrm{sb}}d^2)$, where $A_{\mathrm{sb}}$ is the area of sub-boundary per unit volume and $\gamma_{\mathrm{sb}}$ is the average sub-boundary energy.
If the average sub-boundary edge length is $L$, we can assume that $A_{\mathrm{sb}} = 3/L$ \cite{humphreys2017recrystallization} to get $P_{\mathrm{sb}} = \gamma_{\mathrm{sb}}f_{\mathrm{V}}L/d^2$.
By expressing the dimensionless average dispersoid volume fraction $f_{\mathrm{V}}$ as $f_{\mathrm{L}}d/2$ per texture component, where $f_{\mathrm{L}}$ is the number of dispersoids per boundary length given in Fig. \ref{fig:dispersoids-at-gb-per-comp} (a), and by approximating $L$ by $D_{\mathrm{A}}$, we arrive at a modified expression for the Smith-Zener drag at sub-boundaries per texture component:

\begin{equation}
  P_{\mathrm{sb}}' = \frac{\gamma_{\mathrm{sb}} f_{\mathrm{L}} D_{\mathrm{A}}}{2d} \propto \frac{f_{\mathrm{L}} D_{\mathrm{A}}}{d},
  \label{eq:smith-zener-drag}
\end{equation}

\noindent where the possible differences in $\gamma_{\mathrm{sb}}$ are neglected.
$P_{\mathrm{sb}}'$ per texture component is plotted in Fig. \ref{fig:smith-zener-drag}, with the confidence intervals calculated from error propagation of the confidence intervals of $f_{\mathrm{L}}$, $D_{\mathrm{A}}$ and $d$.
Prior to recovery, all recrystallization subgrains experience a similar drag from dispersoids.
The variations in $P_{\mathrm{sb}}'$ in the as deformed state are mostly due to variations in $D_{\mathrm{A}}$, shown in Fig. \ref{fig:subgrain-stats} (c).
After recovery, Cube sub-boundaries experience the highest drag, while the drag on P and CubeND sub-boundaries increase only slightly, even though all these subgrains have grown in size.
After partial recrystallization, the drag on Cube sub-boundaries remains significantly higher, while the drag on all other sub-boundaries is comparable.

\subsection{Effects of dispersoids on recrystallized grains}

Huang \textit{et al.} \cite{huang2017controlling} reported a recrystallized grain size $D = $ \SI{199.6}{\micro\meter} after non-isothermal annealing of the present alloy to \SI{325}{\celsius} followed by holding at that temperature for \SI{e5}{\second}.
The present experimental approach requires a relatively high BSE image resolution ($\sim$\SI{0.026}{\micro\meter}) and may thus not be the best way to acquire correlated datasets of dispersoids and such large recrystallized grains with statistical significance.
Nonetheless, there are in total 17 recrystallized grains in the correlated datasets from the partly recrystallized state at \SI{325}{\celsius}.
All texture components except for B are represented.
The density $f_{\mathrm{L}}$ and average size $d$ of dispersoids at the boundaries of these recrystallized grains per texture component are presented in Table \ref{tab:dispersoids-at-gb-rx}.
The relatively broad 95\% confidence intervals of the densities reflect the limited number of recrystallized grains analyzed.
However, these results indicate that rapidly growing Cube and CubeND grains experience a higher pinning pressure than rapidly growing P grains during recrystallization, supporting similar observations made by Huang \textit{et al.} \cite{huang2017controlling}.
There is no significant difference in the average dispersoid size on the boundaries.

\begin{table*}[htb]
  \centering
  \caption{
  Dispersoid density $f_{\mathrm{L}}$ and average dispersoid size $d$ with 95\% confidence intervals at the boundaries of 17 recrystallized grains in the partly recrystallized state at \SI{325}{\celsius} of the cold-rolled and non-isothermally annealed Al-Mn alloy.
  }
  \begin{tabular}{l c c c c c}
    \toprule
    Texture comp. & Random & C, S & Cube & CubeND & P \\
    \midrule
    $f_{\mathrm{L}}$ [\si{\per\micro\meter}] & \SI{0.31(4)}{} & \SI{0.44(5)}{} & \SI{0.98(22)}{} & \SI{0.78(12)}{} & \SI{0.13(5)}{} \\
    $d$ [\si{\micro\meter}] & \SI{0.066(3)}{} & \SI{0.064(2)}{} & \SI{0.064(6)}{} & \SI{0.056(3)}{} & \SI{0.061(4)}{} \\
    \bottomrule
  \end{tabular}
  \label{tab:dispersoids-at-gb-rx}
\end{table*}

\begin{figure}[htbp]
  \centering
  \includegraphics[width=0.3\columnwidth]{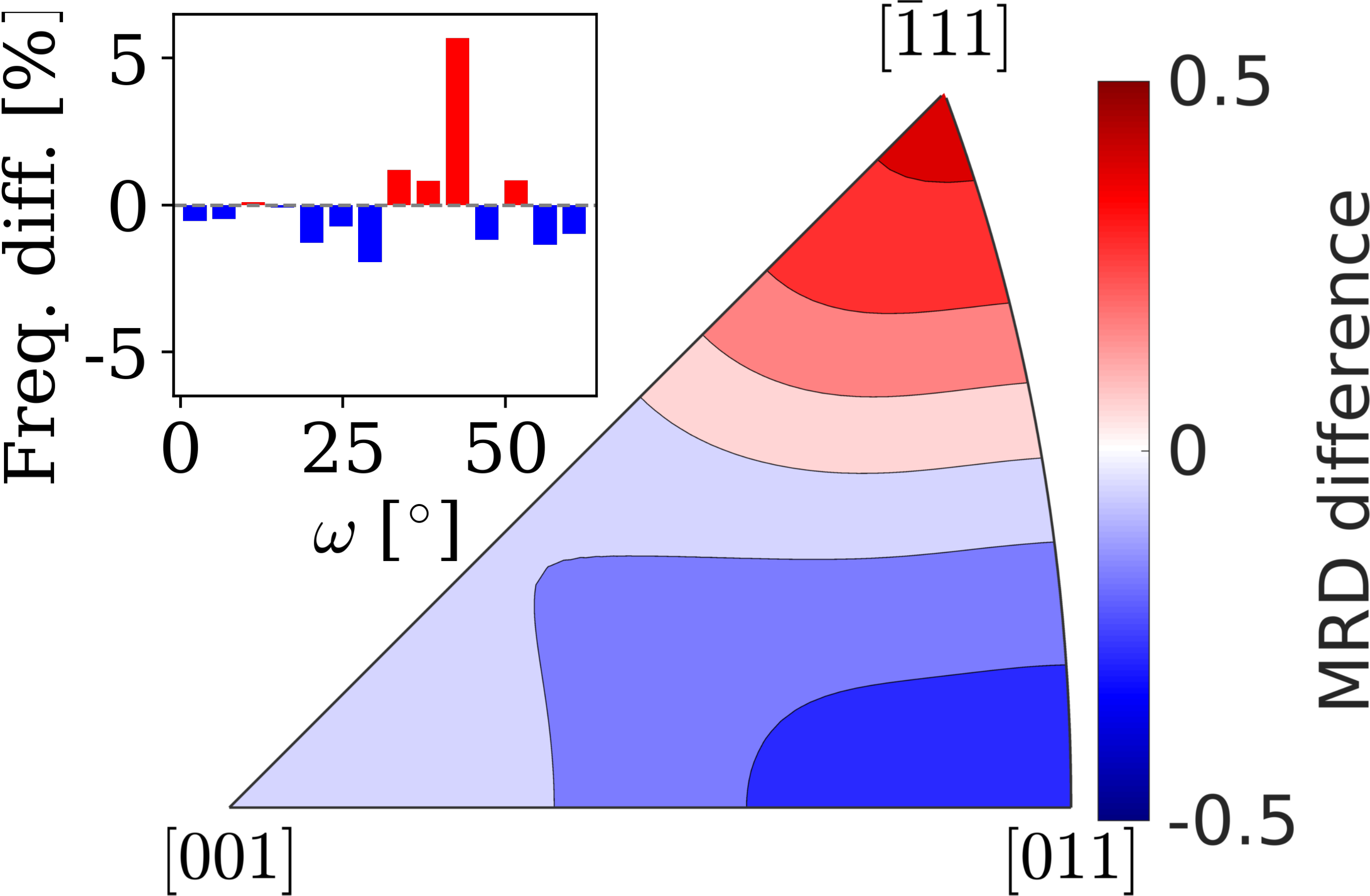}
  \caption{
  Dependence of dispersoid densities $f_{\mathrm{L}}$ on misorientation axis and angle of recrystallized grain boundaries, with higher and lower densities above (red) and below (blue) zero, respectively.
  Relative densities are calculated for the material after partial recrystallization at \SI{325}{\celsius} of the cold-rolled and non-isothermally annealed Al-Mn alloy.
  }
  \label{fig:dispersoids-misorientations-rx}
\end{figure}

The variations in relative dispersoid densities on boundaries of recrystallized grains with misorientation axis and angle, as shown in Fig. \ref{fig:dispersoids-misorientations} for sub-boundaries, are shown in Fig. \ref{fig:dispersoids-misorientations-rx}.
The densities vary much more on recrystallized grain boundaries than on the sub-boundaries.
There is a clearly higher density of dispersoids on 35-45\si{\degree}-boundaries and $\left<111\right>$-boundaries which is not seen on the sub-boundaries at the same partly recrystallized condition [Fig. \ref{fig:dispersoids-misorientations} (d)].
That $\left<111\right>$-boundaries of subgrains after recovery and of recrystallized grains after partial recrystallization have higher dispersoid densities than other boundaries are strong indications that $\left<111\right>$-boundaries are more mobile and thus sweep more dispersoids during subgrain growth and recrystallization, respectively.
Note that the 35-45\si{\degree}-boundaries and $\left<111\right>$-boundaries are not necessarily $\Sigma$7-boundaries.
The boundaries of more recrystallized grains must be incorporated in the analysis to allow such a classification of boundaries into different types.

\section{Discussion}

It is clear from the presented results that particles affect the nucleation and growth of Cube, CubeND and P (sub)grains differently during recovery and recrystallization.
In the following, a brief summary of the nucleation and growth mechanisms of these textures in cold-rolled Al-Mn alloys is given, followed by a discussion of the present results based on a competition between these mechanisms in the presence of concurrent precipitation.

Recrystallization textures in Al alloys are generally characterized by preferential nucleation of subgrains of certain orientations, followed by preferential growth of some of these subgrains to become recrystallized grains \cite{engler1996nucleation,doherty1997current,humphreys2017recrystallization}.
The main nucleation sites of Cube orientations are band-like structures already present in the as deformed microstructure and retained HAGBs \cite{hjelen1991on,engler1996formation}.
The Cube nuclei have been observed to recover faster than other nuclei, either dynamically during deformation or statically during recovery, thus gaining an often considerable size advantage \cite{bunkholt2019orientation,sukhopar2012investigation}.
In addition to these favorable nucleation conditions, Cube grains have a close to \SI{38}{\degree}$\left<111\right>$ orientation relationship, or $\Sigma$7 in coincident site lattice terminology, to grains of the S orientation, which is, as the C and B orientations, a significant deformation texture component in most cold-rolled Al alloys, and also in the present alloy.
Grains with this orientation relationship to neighboring grains have been observed to grow faster than grains with general HAGBs \cite{lucke1974orientation,fan2014oriented}.
As for the CubeND and P orientations, as mentioned above in \S\ref{sec:effects-constituent-particles-subgrains}, a slight preference for these orientations have been observed among the subgrains growing out of deformation zones around larger particles, i.e. PSN.
This preference of CubeND and P nuclei is explained by their close to $\Sigma7$-orientation relationships to C grains, which have been observed experimentally \cite{engler1996formation,huang2017controlling}.

There are many reports on the development of strong CubeND and P recrystallization textures in supersaturated Al-Mn alloys after significant concurrent precipitation \cite{tangen2010effect,zhao2016orientation,huang2014influence,huang2014microstructural,huang2017controlling}.
The preference of these textures over the Cube texture is mainly attributed to two mechanisms: (1) an increased suppression of nucleation from Cube-bands and HAGBs by existing and concurrently precipitated dispersoids, effectively promoting the effect of PSN, and (2) an increased growth advantage of the CubeND and P textures due to a reduced solute drag and Smith-Zener pinning effect from dispersoids because, presumably, a larger fraction of their boundaries are of $\Sigma7$-type.

The results presented in this work clearly indicate an increased suppression of Cube nuclei.
The initial size advantage of Cube subgrains during recovery seen in Fig. \ref{fig:subgrain-stats} (c) indicates a faster recovery, which results in an increased Smith-Zener drag $P_{\mathrm{sb}}'$ due to the high density of dispersoids precipitated prior to recrystallization (see Fig. \ref{fig:particle-stats} (a, b) and e.g. Refs. \cite{huang2014influence,huang2017controlling}).
There are also relatively more and smaller dispersoids in the microstructure away from constituent particles where Cube nucleation sites are typically located.
This increased pinning effectively hinders the initial size advantage of Cube subgrains to become significant, and thus the domination of the Cube texture.
The importance of the size advantage for the Cube texture was discussed by Vatne \textit{et al.} in a supersaturated AA3103 alloy \cite{vatne1997influence}.
Their alloy had a similar microchemistry to the present alloy, with a strong potential for PSN, due to a considerable amount of large particles, and a significant concurrent precipitation reaction.
In contrast to the present recrystallized texture, they observed a texture dominated by Cube, which was explained by the Cube subgrains reaching an overcritical size sooner than PSN subgrains, and thus not experiencing as strong a Smith-Zener drag as the latter subgrains.
That an orientation dependent Smith-Zener drag in these similar alloys results in very different recrystallization textures illustrates the delicate balance between nucleation and growth mechanisms during recovery and recrystallization with concurrent precipitation.

The promoted effect of other nucleation mechanisms due to the weakened nucleation of the Cube texture, like PSN, is a strong contributing factor to the increased P and CubeND textures.
The P and CubeND textures also have a frequency advantage over Cube in the as deformed state, especially P, as seen in Fig. \ref{fig:subgrain-stats} (a, b).
This observation is made possible due to the relatively robust indexing of small subgrains in highly deformed Al-alloys using dictionary indexing \cite{singh2018high}.
The frequency advantage of P and CubeND among the subgrains in deformation zones around larger particles is even greater [see Fig. \ref{fig:subgrains-at-constituent-particles} (a)].
The reduced Smith-Zener pinning potential in the deformation zones due to fewer and larger dispersoids in these zones strengthens this frequency advantage.
The P texture has a slight frequency advantage over the CubeND texture [Fig. \ref{fig:subgrain-stats} (a, b)], P subgrains seem to grow faster than CubeND subgrains during recovery [Fig. \ref{fig:subgrain-stats} (c)], while P subgrains at constituent particles seem to grow faster than CubeND subgrains during recrystallization [Fig. \ref{fig:subgrains-at-constituent-particles} (b)].
Apart from these differences, the present work indicates that effects from PSN and from the modified Smith-Zener drag $P'_{\mathrm{sb}}$ vary little for P and CubeND subgrains.

The only substantial difference between the P and CubeND textures is the much lower dispersoid density $f_{\mathrm{L}}$ on boundaries of recrystallized P grains compared to that on CubeND grain boundaries in the partly recrystallized state [see Table \ref{tab:dispersoids-at-gb-rx}].
The reason for this difference is not clear from the present data.
However, such a substantial difference could explain the observed dominance of P compared to the CubeND texture.
It should be noted that this observation is based on limited statistical evidence of four P and two CubeND recrystallized grains.
A natural continuation of the present work is to investigate this difference in $f_{\mathrm{L}}$ in the partly recrystallized state further to achieve better statistics.

We have in this work demonstrated the application of a novel correlated procedure, recently introduced in a related paper by the authors \cite{aanes2022correlated}, to directly correlate (sub)grains and particles during recovery and recrystallization in an Al-Mn alloy.
The methodology combines (sub)grain data from EBSD and particle data from BSE images, enabling a new correlated quantitative analysis of (sub)grains in deformation zones surrounding constituent particles and dispersoids located at (sub)grain boundaries.
It is demonstrated that this methodology enables a better understanding of the mechanisms behind the orientation dependent nucleation and growth behavior during recovery and recrystallization in Al-Mn alloys in conditions of strong concurrent precipitation, and in particular the weak Cube texture compared to the stronger P and CubeND textures.


\section{Concluding remarks}

Multimodal datasets comprising direct spatial correlation of constituent particles and dispersoids with the microstructure in terms of (sub)grains and (sub-)boundaries were analyzed based on BSE imaging and EBSD.
The presented results suggest the following causes for the dominating P and CubeND recrystallization textures in a cold-rolled and non-isothermally annealed Al-Mn alloy with concurrent precipitation:

\begin{itemize}
  \item The often observed size advantage of Cube subgrains is not achieved due to an increased Smith-Zener drag on their boundaries after recovery and partial recrystallization, thus promoting the effect of PSN.
  \item Dispersoids in deformation zones around large particles are fewer and larger compared to elsewhere in the microstructure, providing a reduced Smith-Zener drag for PSN nuclei.
  \item The only advantages of the P texture compared to the CubeND texture are a frequency advantage both in deformation zones around large particles and elsewhere in the microstructure, and a slight size advantage due to faster subgrain growth.
  \item Indications of a much lower dispersoid density on the boundaries of recrystallized P grains compared to those on CubeND and Cube boundaries were observed, although investigations with better statistics are required to further substantiate this result.
\end{itemize}


\section*{Data and code availability}

The raw EBSD datasets and BSE images required to reproduce these findings are available from Zenodo at \url{https://doi.org/10.5281/zenodo.7383087} \cite{aanes2022electron2}.
All softwares used for analysis of the data are freely available, apart from \textit{MTEX} which requires a \textit{MATLAB} license.
\textit{Jupyter} notebooks and \textit{MTEX} and \textit{ImageJ} scripts for most data processing steps are available from the following \textit{GitHub} repository: \url{https://github.com/hakonanes/p-texture-al-mn-alloys}.


\section*{Declaration of Competing Interest}

The authors declare that they have no known competing financial interests or personal relationships that could have appeared to influence the work reported in this paper.


\section*{Acknowledgements}

HWÅ acknowledges NTNU for financial support through the NTNU Aluminium Product Innovation Center (NAPIC).
The authors would like to thank Hydro Aluminium for supplying the material.


\bibliographystyle{elsarticle-num}
\bibliography{library_sanitized}

\begin{thebibliography}{10}
\expandafter\ifx\csname url\endcsname\relax
  \def\url#1{\texttt{#1}}\fi
\expandafter\ifx\csname urlprefix\endcsname\relax\def\urlprefix{URL }\fi
\expandafter\ifx\csname href\endcsname\relax
  \def\href#1#2{#2} \def\path#1{#1}\fi

\bibitem{hirsch2013superior}
J.~Hirsch, T.~Al-Samman, {Superior light metals by texture engineering:
  Optimized aluminum and magnesium alloys for automotive applications}, Acta
  Materialia 61~(3) (2013) 818–843.
\newblock \href {https://doi.org/10.1016/j.actamat.2012.10.044}
  {\path{doi:10.1016/j.actamat.2012.10.044}}.

\bibitem{humphreys1977nucleation}
F.~J. Humphreys, {The nucleation of recrystallization at second phase particles
  in deformed aluminium}, Acta Metallurgica 25~(11) (1977) 1323–1344.
\newblock \href {https://doi.org/10.1016/0001-6160(77)90109-2}
  {\path{doi:10.1016/0001-6160(77)90109-2}}.

\bibitem{engler1996nucleation}
O.~Engler, {Nucleation and growth during recrystallisation of aluminium alloys
  investigated by local texture analysis}, Materials Science and Technology
  12~(10) (1996) 859–872.
\newblock \href {https://doi.org/10.1179/026708396790122206}
  {\path{doi:10.1179/026708396790122206}}.

\bibitem{rohrer2010introduction}
G.~S. Rohrer, {“Introduction to Grains, Phases, and Interfaces—an
  Interpretation of Microstructure,” Trans. AIME, 1948, vol. 175, pp.
  15–51, by CS Smith}, Metallurgical and Materials Transactions B 41~(3)
  (2010) 457–494.
\newblock \href {https://doi.org/10.1007/s11663-010-9364-6}
  {\path{doi:10.1007/s11663-010-9364-6}}.

\bibitem{humphreys2017recrystallization}
F.~J. Humphreys, G.~S. Rohrer, A.~D. Rollett, {Recrystallization and Related
  Annealing Phenomena}, 3rd Edition, Elsevier, Oxford, 2017.

\bibitem{huang2018doubleedge}
K.~Huang, K.~Marthinsen, Q.~Zhao, R.~E. Logé, {The double-edge effect of
  second-phase particles on the recrystallization behaviour and associated
  mechanical properties of metallic materials}, Progress in Materials Science
  92 (2018) 284–359.
\newblock \href {https://doi.org/10.1016/j.pmatsci.2017.10.004}
  {\path{doi:10.1016/j.pmatsci.2017.10.004}}.

\bibitem{vatne1997influence}
H.~Vatne, O.~Engler, E.~Nes, {Influence of particles on recrystallisation
  textures and microstructures of aluminium alloy 3103}, Materials Science and
  Technology 13~(2) (1997) 93–102.
\newblock \href {https://doi.org/10.1179/mst.1997.13.2.93}
  {\path{doi:10.1179/mst.1997.13.2.93}}.

\bibitem{somerday2003recrystallisation}
M.~Somerday, F.~Humphreys, {Recrystallisation behaviour of supersaturated
  Al–Mn alloys Part 1–Al–1.3 wt-\% Mn}, Materials Science and Technology
  19~(1) (2003) 20–29.
\newblock \href {https://doi.org/10.1179/026708303225008626}
  {\path{doi:10.1179/026708303225008626}}.

\bibitem{tangen2010effect}
S.~Tangen, K.~Sjølstad, T.~Furu, E.~Nes, {Effect of concurrent precipitation
  on recrystallization and evolution of the P-texture component in a commercial
  Al-Mn alloy}, Metallurgical and Materials Transactions A: Physical Metallurgy
  and Materials Science 41~(11) (2010) 2970–2983.
\newblock \href {https://doi.org/10.1007/s11661-010-0265-8}
  {\path{doi:10.1007/s11661-010-0265-8}}.

\bibitem{huang2014influence}
K.~Huang, N.~Wang, Y.~Li, K.~Marthinsen, {The influence of microchemistry on
  the softening behaviour of two cold-rolled Al–Mn–Fe–Si alloys},
  Materials Science and Engineering: A 601 (2014) 86–96.
\newblock \href {https://doi.org/10.1016/j.msea.2014.02.037}
  {\path{doi:10.1016/j.msea.2014.02.037}}.

\bibitem{zhao2016orientation}
Q.~Zhao, K.~Huang, Y.~Li, K.~Marthinsen, {Orientation Preference of
  Recrystallization in Supersaturated Aluminum Alloys Influenced by Concurrent
  Precipitation}, Metallurgical and Materials Transactions A: Physical
  Metallurgy and Materials Science 47~(3) (2016) 1378–1388.
\newblock \href {https://doi.org/10.1007/s11661-015-3314-5}
  {\path{doi:10.1007/s11661-015-3314-5}}.

\bibitem{huang2017controlling}
K.~Huang, K.~Zhang, K.~Marthinsen, R.~E. Logé, {Controlling grain structure
  and texture in Al-Mn from the competition between precipitation and
  recrystallization}, Acta Materialia 141 (2017) 360–373.
\newblock \href {https://doi.org/10.1016/j.actamat.2017.09.032}
  {\path{doi:10.1016/j.actamat.2017.09.032}}.

\bibitem{aanes2022correlated}
H.~W. Ånes, A.~T.~J. {van Helvoort}, K.~Marthinsen, {Correlated subgrain and
  particle analysis of a recovered Al-Mn alloy by directly combining EBSD and
  backscatter electron imaging}, Materials Characterization (2022).
\newblock \href {https://doi.org/10.1016/j.matchar.2022.112228}
  {\path{doi:10.1016/j.matchar.2022.112228}}.

\bibitem{hielscher2008novel}
R.~Hielscher, H.~Schaeben, {A novel pole figure inversion method: Specification
  of the MTEX algorithm}, Journal of Applied Crystallography 41~(6) (2008)
  1024–1037.
\newblock \href {https://doi.org/10.1107/S0021889808030112}
  {\path{doi:10.1107/S0021889808030112}}.

\bibitem{schneider2012nih}
C.~A. Schneider, W.~S. Rasband, K.~W. Eliceiri, {NIH Image to ImageJ: 25 years
  of image analysis}, Nature Methods 9~(7) (2012) 671--675.
\newblock \href {https://doi.org/10.1038/nmeth.2089}
  {\path{doi:10.1038/nmeth.2089}}.

\bibitem{chen2015dictionary}
Y.~H. Chen, S.~U. Park, D.~Wei, G.~Newstadt, M.~A. Jackson, J.~P. Simmons,
  M.~{De Graef}, A.~O. Hero, {A Dictionary Approach to Electron Backscatter
  Diffraction Indexing}, Microscopy and Microanalysis 21~(3) (2015) 739–752.
\newblock \href {https://doi.org/10.1017/S1431927615000756}
  {\path{doi:10.1017/S1431927615000756}}.

\bibitem{pena2017electron}
F.~de~la Peña, T.~Ostasevicius, V.~T. Fauske, P.~Burdet, P.~Jokubauskas,
  M.~Nord, M.~Sarahan, E.~Prestat, D.~N. Johnstone, J.~Taillon, et~al.,
  {Electron microscopy (Big and Small) data analysis with the open source
  software package HyperSpy}, Microscopy and Microanalysis 23~(S1) (2017)
  214–215.
\newblock \href {https://doi.org/10.1017/S1431927617001751}
  {\path{doi:10.1017/S1431927617001751}}.

\bibitem{kikuchipy_0_6_1}
H.~W. Ånes, L.~Lervik, O.~Natlandsmyr, T.~Bergh, Z.~Xu, E.~Prestat,
  pyxem/kikuchipy: kikuchipy 0.6.1 (Jun. 2022).
\newblock \href {https://doi.org/10.5281/zenodo.6655562}
  {\path{doi:10.5281/zenodo.6655562}}.

\bibitem{callahan2013dynamical}
P.~G. Callahan, M.~{De Graef}, {Dynamical Electron Backscatter Diffraction
  Patterns. Part I: Pattern Simulations}, Microscopy and Microanalysis 19
  (2013) 1255–1265.
\newblock \href {https://doi.org/10.1017/S1431927613001840}
  {\path{doi:10.1017/S1431927613001840}}.

\bibitem{pyebsdindex}
D.~Rowenhorst, H.~W. Ånes, {{USNavalResearchLaboratory/PyEBSDIndex:
  PyEBSDIndex}: 0.1.0} (2022).

\bibitem{bachmann2011grain}
F.~Bachmann, R.~Hielscher, H.~Schaeben, {Grain detection from 2d and 3d EBSD
  data-Specification of the MTEX algorithm}, Ultramicroscopy 111~(12) (2011)
  1720–1733.
\newblock \href {https://doi.org/10.1016/j.ultramic.2011.08.002}
  {\path{doi:10.1016/j.ultramic.2011.08.002}}.

\bibitem{singh2018high}
S.~Singh, Y.~Guo, B.~Winiarski, T.~L. Burnett, P.~J. Withers, M.~{De Graef},
  {High resolution low kV EBSD of heavily deformed and nanocrystalline
  Aluminium by dictionary-based indexing}, Scientific reports 8 (2018).
\newblock \href {https://doi.org/10.1038/s41598-018-29315-8}
  {\path{doi:10.1038/s41598-018-29315-8}}.

\bibitem{furrer1979recrystallization}
P.~Furrer, G.~Hausch, {Recrystallization behaviour of commercial Al-1\%Mn
  alloy}, Metal Science 13~(3-4) (1979) 155--162.
\newblock \href {https://doi.org/10.1179/msc.1979.13.3-4.155}
  {\path{doi:10.1179/msc.1979.13.3-4.155}}.

\bibitem{zhang2012three}
Y.~Zhang, D.~J. Jensen, Y.~Zhang, F.~Lin, Z.~Zhang, Q.~Liu, {Three-dimensional
  investigation of recrystallization nucleation in a particle-containing Al
  alloy}, Scripta Materialia 67~(4) (2012) 320–323.
\newblock \href {https://doi.org/10.1016/j.scriptamat.2012.05.006}
  {\path{doi:10.1016/j.scriptamat.2012.05.006}}.

\bibitem{engler1996formation}
O.~Engler, P.~Yang, X.~W. Kong, {On the formation of recrystallization textures
  in binary A1-1.3{\%} Mn investigated by means of local texture analysis},
  Acta Materialia 44~(8) (1996) 3349–3369.
\newblock \href {https://doi.org/10.1016/1359-6454(95)00416-5}
  {\path{doi:10.1016/1359-6454(95)00416-5}}.

\bibitem{engler1995texture}
O.~Engler, J.~Hirsch, K.~L{\"u}cke, {Texture development in Al-1.8 wt\% Cu
  depending on the precipitation state—II. Recrystallization textures}, Acta
  Metallurgica Et Materialia 43~(1) (1995) 121--138.
\newblock \href {https://doi.org/10.1016/0956-7151(95)90268-6}
  {\path{doi:10.1016/0956-7151(95)90268-6}}.

\bibitem{schafer2011origin}
C.~Schäfer, G.~Gottstein, The origin and development of the
  {P}\{011\}$\langle$111$\rangle$ orientation during recrystallization of
  particle-containing alloys, International Journal of Materials Research
  102~(9) (2011) 1106–1114.
\newblock \href {https://doi.org/10.3139/146.110567}
  {\path{doi:10.3139/146.110567}}.

\bibitem{lucke1974orientation}
K.~L{\"u}cke, The orientation dependence of grain boundary motion and the
  formation of recrystallization textures, Canadian Metallurgical Quarterly
  13~(1) (1974) 261--274.
\newblock \href {https://doi.org/10.1179/cmq.1974.13.1.261}
  {\path{doi:10.1179/cmq.1974.13.1.261}}.

\bibitem{yang2001measuring}
C.-C. Yang, A.~Rollett, W.~Mullins, Measuring relative grain boundary energies
  and mobilities in an aluminum foil from triple junction geometry, Scripta
  Materialia 44~(12) (2001) 2735--2740.
\newblock \href {https://doi.org/10.1016/S1359-6462(01)00960-5}
  {\path{doi:10.1016/S1359-6462(01)00960-5}}.

\bibitem{fan2014oriented}
G.~Fan, Y.~Zhang, J.~H. Driver, D.~J. Jensen, {Oriented growth during
  recrystallization revisited in three dimensions}, Scripta Materialia 72
  (2014) 9–12.
\newblock \href {https://doi.org/10.1016/j.scriptamat.2013.09.031}
  {\path{doi:10.1016/j.scriptamat.2013.09.031}}.

\bibitem{hutchinson1978influence}
W.~Hutchinson, B.~Duggan, {Influence of precipitation on recrystallization and
  texture development in an iron-1.2\% copper alloy}, Metal Science 12~(8)
  (1978) 372–380.
\newblock \href {https://doi.org/10.1179/msc.1978.12.8.372}
  {\path{doi:10.1179/msc.1978.12.8.372}}.

\bibitem{doherty1997current}
R.~D. Doherty, D.~A. Hughes, F.~J. Humphreys, J.~J. Jonas, D.~J. Jensen, M.~E.
  Kassner, W.~E. King, T.~R. McNelley, H.~J. McQueen, A.~D. Rollett, {Current
  issues in recrystallization: a review}, Materials Science and Engineering: A
  238~(2) (1997) 219–274.
\newblock \href {https://doi.org/10.1016/S0921-5093(97)00424-3}
  {\path{doi:10.1016/S0921-5093(97)00424-3}}.

\bibitem{hjelen1991on}
J.~Hjelen, R.~Ørsund, E.~Nes, {On the Origin of Recrystallization Textures in
  Aluminium}, Acta Metallurgica Et Materialia 39~(7) (1991) 1377–1404.
\newblock \href {https://doi.org/10.1016/0956-7151(91)90225-P}
  {\path{doi:10.1016/0956-7151(91)90225-P}}.

\bibitem{bunkholt2019orientation}
S.~Bunkholt, E.~Nes, K.~Marthinsen, {Orientation independent and dependent
  subgrain growth during iso-thermal annealing of high-purity and commercial
  purity aluminium}, Metals 9~(10) (2019) 1032.
\newblock \href {https://doi.org/10.3390/met9101032}
  {\path{doi:10.3390/met9101032}}.

\bibitem{sukhopar2012investigation}
O.~Sukhopar, M.~Kuzmina, G.~Gottstein, {Investigation of the substructure
  evolution within grains of different orientations during recrystallization in
  a commercial Al alloy}, in: {Advanced Materials Research}, Vol. 409, Trans
  Tech Publ, 2012, p. 71–76.
\newblock \href {https://doi.org/10.4028/www.scientific.net/AMR.409.71}
  {\path{doi:10.4028/www.scientific.net/AMR.409.71}}.

\bibitem{huang2014microstructural}
K.~Huang, Y.~J. Li, K.~Marthinsen, Microstructural evolution during isothermal
  annealing of a cold-rolled al-mn-fe-si alloy with different microchemistry
  states, in: Materials Science Forum, Vol. 794, Trans Tech Publ, 2014, pp.
  1163--1168.
\newblock \href {https://doi.org/10.4028/www.scientific.net/MSF.794-796.1163}
  {\path{doi:10.4028/www.scientific.net/MSF.794-796.1163}}.

\bibitem{aanes2022electron2}
H.~W. Ånes, A.~T.~J. van Helvoort, K.~Marthinsen, {Electron backscatter
  diffraction data and backscatter electron images from four conditions from a
  cold-rolled and annealed Al-Mn alloy} (2022).
\newblock \href {https://doi.org/10.5281/zenodo.7383087}
  {\path{doi:10.5281/zenodo.7383087}}.

\end{thebibliography}


\section*{Supplementary material}

\renewcommand{\thefigure}{S\arabic{figure}}

\setcounter{figure}{0}

\begin{figure*}[htbp]
  \centering
  \includegraphics[width=\textwidth]{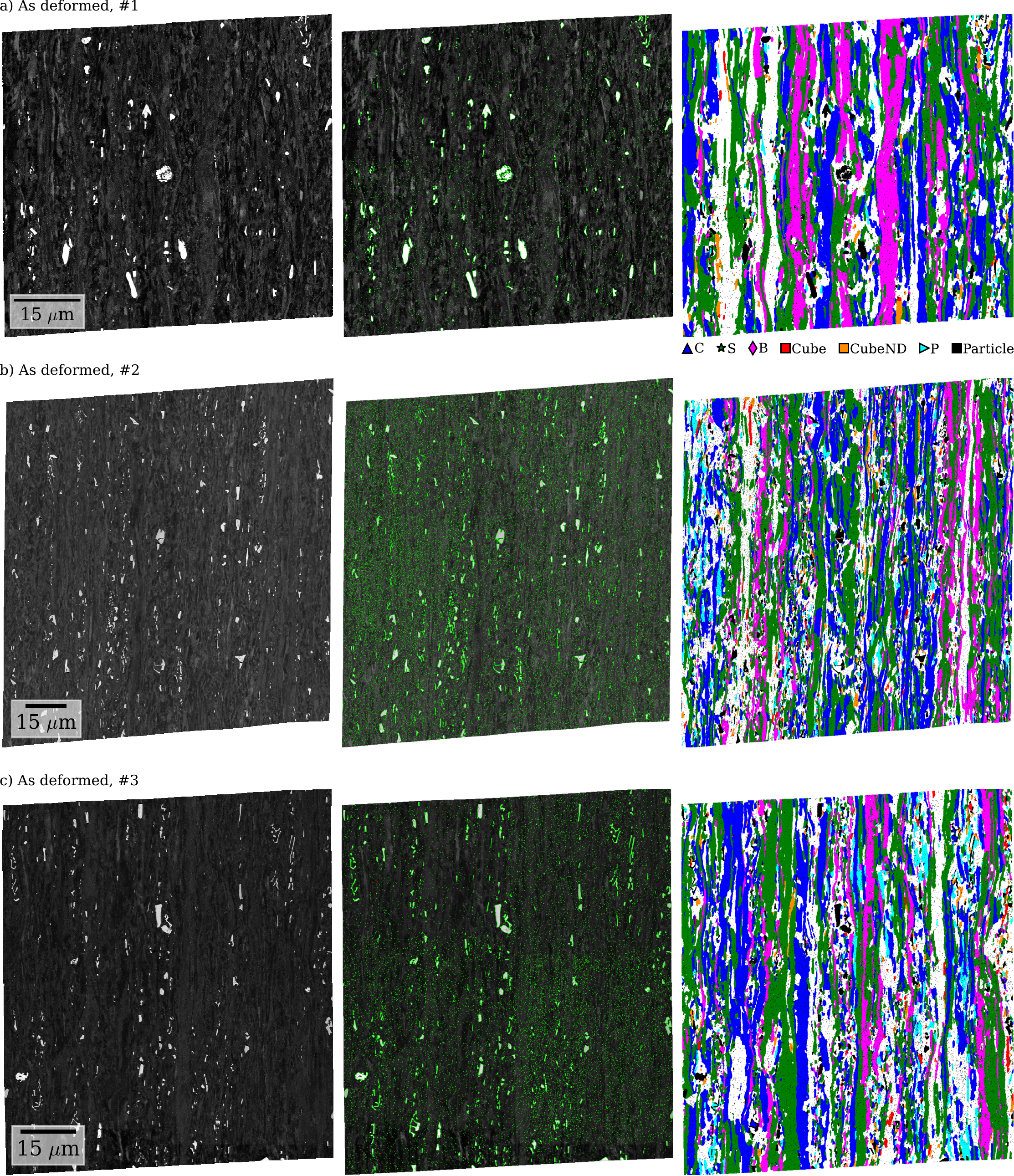}
  \caption{
  Multimodal datasets from the as deformed state.
  BSE images (left column), BSE images with correctly detected (green) and incorrectly detected (red) particles (center column) and the grains from EBSD colored according to the ideal texture components (right column).
  The legend for the texture components is given in (a).
  Black pixels in the EBSD maps correspond to detected particles in the BSE images.
  }
  \label{fig:maps-0}
\end{figure*}

\begin{figure*}[htbp]
  \centering
  \includegraphics[width=\textwidth]{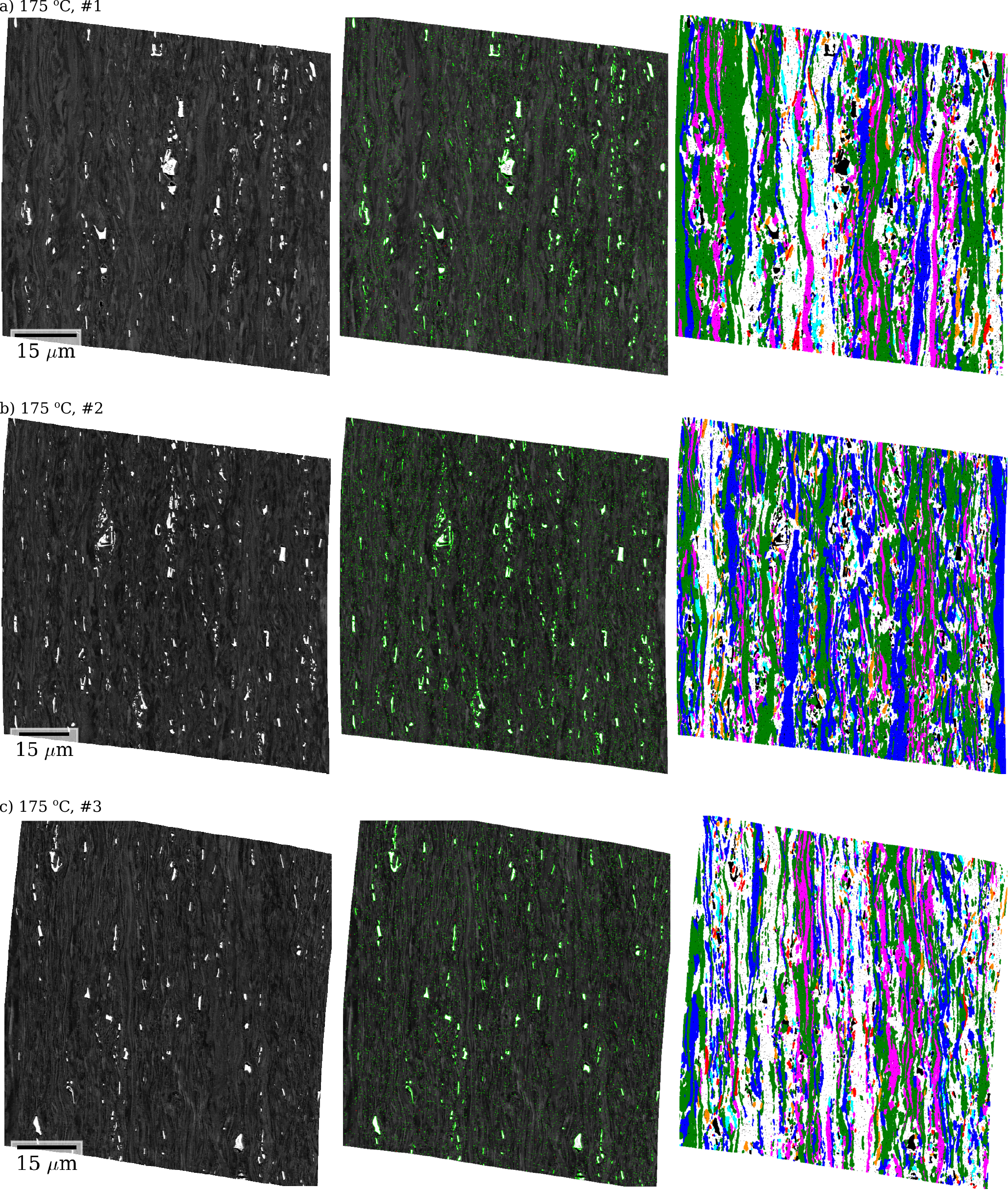}
  \caption{
  Multimodal dataset from the condition at \SI{175}{\celsius}.
  BSE images (left column), BSE images with correctly detected (green) and incorrectly detected (red) particles (center column) and the grains from EBSD colored according to the ideal texture components (right column).
  The legend for the texture components is given in Fig. \ref{fig:maps-0} (a).
  Black pixels in the EBSD maps correspond to detected particles in the BSE images.
  }
  \label{fig:maps-175c}
\end{figure*}

\begin{figure*}[htbp]
  \centering
  \includegraphics[width=\textwidth]{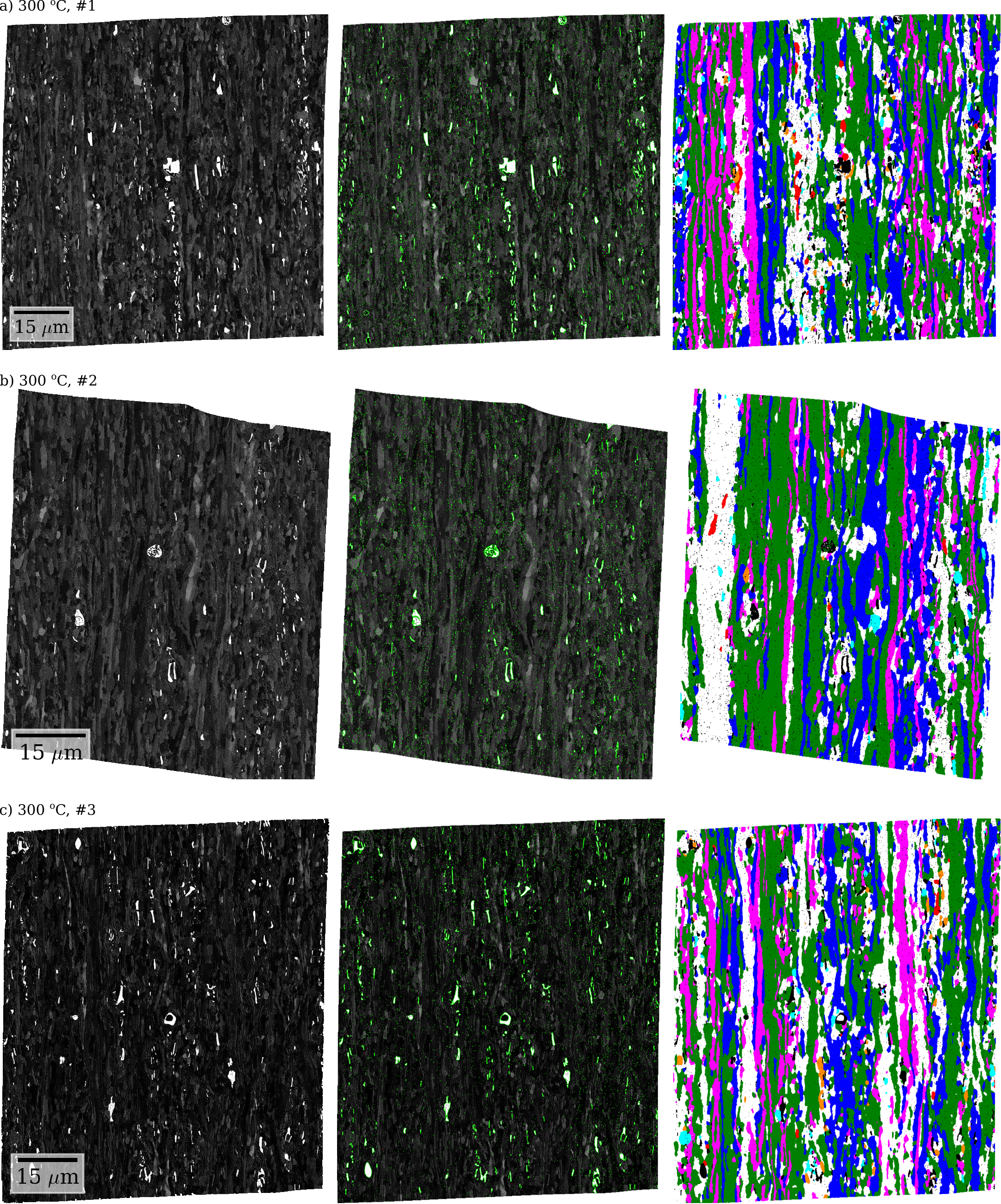}
  \caption{
  Multimodal dataset from the condition at \SI{300}{\celsius}.
  BSE images (left column), BSE images with correctly detected (green) and incorrectly detected (red) particles (center column) and the grains from EBSD colored according to the ideal texture components (right column).
  The legend for the texture components is given in Fig. \ref{fig:maps-0} (a).
  Black pixels in the EBSD maps correspond to detected particles in the BSE images.
  }
  \label{fig:maps-300c}
\end{figure*}

\begin{figure*}[htbp]
  \centering
  \includegraphics[width=\textwidth]{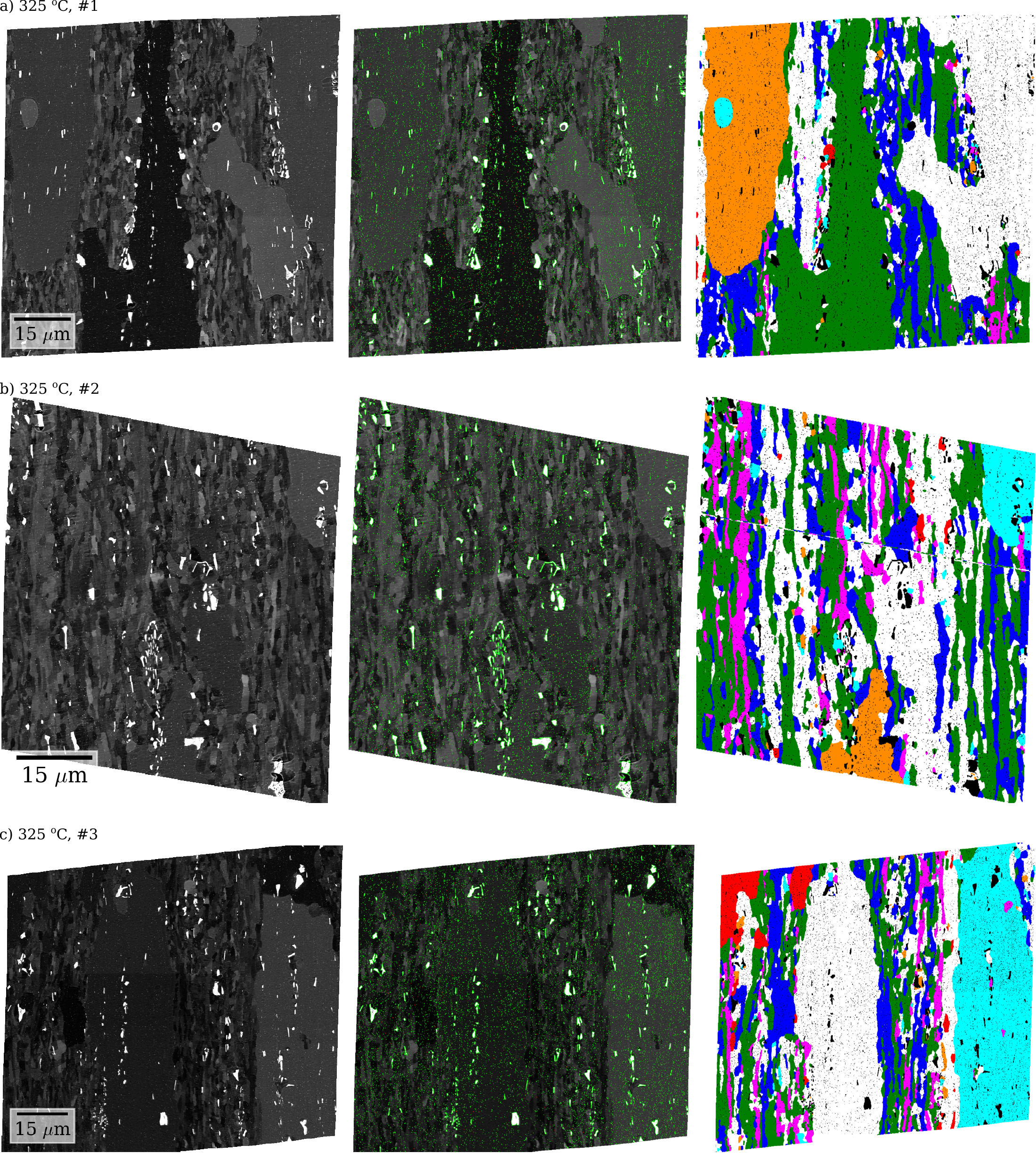}
  \caption{
  Multimodal dataset from the condition at \SI{325}{\celsius}.
  BSE images (left column), BSE images with correctly detected (green) and incorrectly detected (red) particles (center column) and the grains from EBSD colored according to the ideal texture components (right column).
  The legend for the texture components is given in Fig. \ref{fig:maps-0} (a).
  Black pixels in the EBSD maps correspond to detected particles in the BSE images.
  }
  \label{fig:maps-325c}
\end{figure*}

\begin{figure*}[htbp]
  \centering
  \includegraphics[width=\textwidth]{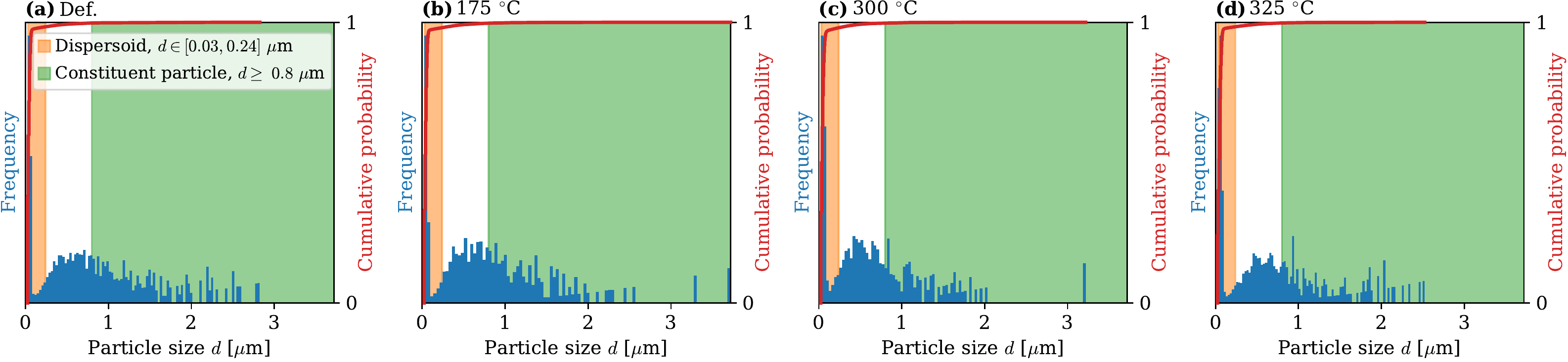}
  \caption{
  Area weighted distribution of particle sizes $d$ and the cumulative probability distribution of $d$, in the cold-rolled and non-isothermally annealed Al-Mn alloy at selected temperatures (a-d) from the as deformed (`Def.') to the partly recrystallized state.
  Particles with a size $d \in [0.03, 0.24]$ \si{\micro\meter} are considered dispersoids while those $d \ge$ \SI{0.8}{\micro\meter} are considered constituent particles potentially contributing to PSN.
  }
  \label{fig:particle-size-hist}
\end{figure*}

\begin{figure}[htbp]
  \centering
  \includegraphics[width=0.5\textwidth]{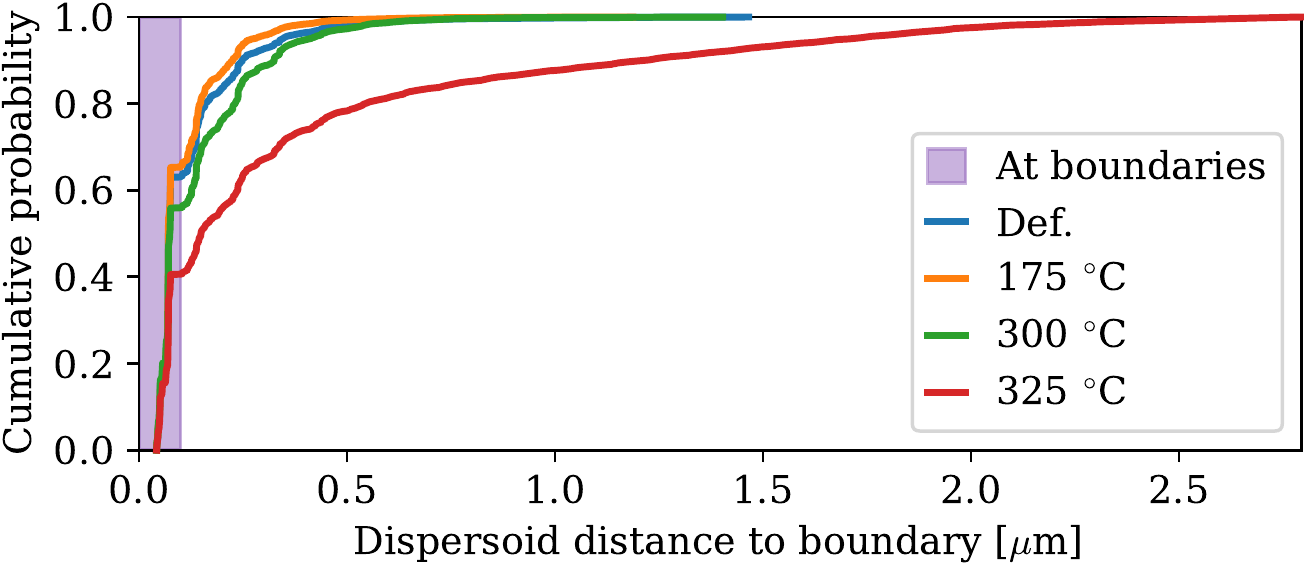}
  \caption{
  Cumulative probability of each dispersoid's shortest distance to any (sub)grain boundary in the cold-rolled and non-isothermally annealed Al-Mn alloy at selected temperatures from the as deformed (`Def.') to the partly recrystallized state.
  Those dispersoids within \SI{0.1}{\micro\meter} (the EBSD step size) of a boundary are considered to be on it.
  }
  \label{fig:dispersoid-distance-to-gb}
\end{figure}


\end{document}